\documentclass[preprint]{aastex}
\usepackage{graphicx}

\begin{document}

\title{The SDSS-III APOGEE Spectral Line List for $H$-band Spectroscopy}

\author{M. Shetrone\altaffilmark{1}, 
D. Bizyaev\altaffilmark{2} \altaffilmark{3} ,
J.E. Lawler\altaffilmark{4},
C. Allende Prieto\altaffilmark{5} \altaffilmark{6},
J.A. Johnson\altaffilmark{7},
V.V. Smith\altaffilmark{8},
K. Cunha\altaffilmark{9}  \altaffilmark{13},
J. Holtzman\altaffilmark{11},
A.E. Garc\'{\i}a P\'erez\altaffilmark{12},
Sz. M{\'e}sz{\'a}ros\altaffilmark{13},
J. Sobeck\altaffilmark{12},
O. Zamora\altaffilmark{5} \altaffilmark{6},
D.A. Garc\'{\i}a-Hern{\'a}ndez\altaffilmark{5} \altaffilmark{6},
D. Souto\altaffilmark{5} \altaffilmark{9},
D. Chojnowski\altaffilmark{2},
L. Koesterke\altaffilmark{14},
S. Majewski\altaffilmark{12},
G. Zasowski\altaffilmark{15}
}

\altaffiltext{1}{University of Texas at Austin, McDonald Observatory, USA}
\altaffiltext{2}{Apache Point Observatory and New Mexico State University, P.O. Box 59, Sunspot, NM, 88349-0059, USA}
\altaffiltext{3}{Sternberg Astronomical Institute, Moscow State University, Moscow, Russia}
\altaffiltext{4}{Department of Physics, University of Wisconsin-Madison, 1150 University Ave., Madison, WI 53706, USA}
\altaffiltext{5}{Instituto de Astrof\'{\i}sica de Canarias, Calle V\'{i}a Lactea s/n, E-38205 La Laguna, Tenerife, Spain}
\altaffiltext{6}{Departamento de Astrof\'{\i}sica, Universidad de La Laguna, E-38206 La Laguna, Tenerife, Spain}
\altaffiltext{7}{Department of Astronomy, The Ohio State University, Columbus, OH 43210, USA}
\altaffiltext{8}{National Optical Astronomy Observatory, 950 North Cherry Avenue, Tucson, AZ 85719, USA}
\altaffiltext{9}{Observat\'{o}rio Nacional, Rua General Jose Cristino, 77, 20921-400 S\~{a}o Crist\'{o}v\~{a}o, Rio de Janeiro, RJ, Brazil}
\altaffiltext{10}{University of Arizona, Tucson, AZ 85719, USA}
\altaffiltext{11}{Department of Astronomy, New Mexico State University, Las Cruces, NM 88003, USA}
\altaffiltext{12}{Department of Astronomy, University of Virginia, P.O. Box 400325, Charlottesville, VA 22904-4325, USA}
\altaffiltext{13}{ELTE Gothard Astrophysical Observatory, H-9704 Szombathely, Szent Imre herceg st. 112, Hungary}
\altaffiltext{14}{The University of Texas at Austin, Texas Advanced Computing Center, USA}
\altaffiltext{15}{Department of Physics and Astronomy, Johns Hopkins University, Baltimore, MD 21218, USA}

\begin{abstract}

We present the $H$-band spectral line lists adopted by the Apache Point Observatory Galactic Evolution Experiment (APOGEE).  
The APOGEE line lists comprise astrophysical, theoretical, and laboratory sources from the literature,
as well as newly evaluated astrophysical oscillator strengths and damping parameters.   
We discuss the construction of the APOGEE line list, which is one of the critical inputs for the APOGEE 
Stellar Parameters and Chemical Abundances Pipeline, 
and present three different versions that have been used at various stages of the project.   
The methodology for the newly calculated astrophysical line lists is reviewed.
The largest of these three line lists contains 134,457 molecular and atomic transitions.
In addition to the format adopted to store the data, 
the line lists are available in MOOG, Synspec and Turbospectrum formats.   The limitations of the line lists 
along with guidance for its use on different spectral types are discussed.    
We also present a list of $H$-band
spectral features that are either poorly represented or completely missing in our line list.   This list is
based on the average of a large number of spectral fit residuals for APOGEE observations spanning a 
wide range of stellar parameters.

\end{abstract}


\section{Introduction}

The Apache Point Observatory Galactic Evolution Experiment (APOGEE) is one of the programs that was carried out on
the Sloan Foundation 2.5-m Telescope \citep{gunn01}  by the third stage of the
Sloan Digital Sky Survey \citep[SDSS-III;][]{Eisenstein11}.
APOGEE obtained high resolution (R $\sim$ 22,500) and high signal-to-noise ratio ($S/N > 100$) spectra 
in the $H$-band (1.51-1.70 $\mu$m) for more than 100,000 cool giant stars 
\citep[see][for more information about targeting]{Zasowski13} 
spanning all components of the Milky Way \citep{Majewski15}.     
Stellar parameters and individual chemical abundances are derived from the combined APOGEE spectra \citep{Nidever15}
with the APOGEE Stellar Parameters and Chemical Abundances Pipeline (ASPCAP), which is described in 
detail in \citet{Garcia2015}.   
ASPCAP uses a grid of synthetic spectra to determine stellar parameters and abundances by finding the 
best match to the observed spectra interpolating within the grid.  
For the public data releases, the synthetic spectra have been calculated
using the spectral synthesis code ASS{$\epsilon$}T \citep{Koesterke09}, 
which itself is based, in part, on the synthesis code Synspec \citep{Hubeny2011}.

In order to run ASPCAP on the first year of APOGEE results 
\citep[DR10;][]{Ahn14}, an initial line list was generated and adopted.   In subsequent years, 
improvements were made and different methodologies were adopted in the line list used for the release
of the full three year data set of APOGEE \citep[DR12;][]{DR12}.   
 The APOGEE internal naming scheme for the DR10 and DR12
line lists are 20110510 and 20131216, respectively.  
In the years between DR10 and DR12,  \citet{GarciaPerez13}, \citet{Smith13} and \citet{Cunha2015}
made use of an intermediate line list,  20120216.     
The APOGEE naming scheme includes the year, month and day label to keep track of changes
made to each line list.   Because these dated names are long and difficult to associate with each data product we
will refer hereafter to  20110510,   20120216 and 20131216 as DR10, INT and DR12 line lists, respectively.
The line list continues to evolve in APOGEE-2, an extension of the project as part of SDSS-IV. 

In this paper we document the methodologies employed by APOGEE to produce the three $H$-band 
line lists that were used by ASPCAP (and several independent analyses) to derive stellar parameters and 
chemical abundances.   
The first section of this paper describes the base line list taken from literature sources, including those derived from 
laboratory, theoretical, and astrophysical sources.   Parameters of the transitions, even if they are well studied, 
have associated uncertainties.  We aspire to improve on the theoretical line parameters by 
comparing synthetic spectra with the observed high-resolution
spectra of two well known stars (the Sun and Arcturus) 
to produce ``astrophysical'' oscillator strength values and damping 
constants.    In the second section, we describe the
code used to derive our astrophysical log($gf$) and damping 
parameters.    
We also describe some issues identified in the INT and DR12 APOGEE line lists and
describe the impact these issues have on the stellar parameters and 
abundances derived from those lists.   
In Section \ref{missing} we detail the stellar features which appear to be missing from our line 
list, based on  synthesis from a large portion of the DR12 stellar library.  
In the last section, we describe the line list formats and review the performance of the line list as described in a 
number of papers in the literature.

\section{Base Line List from Literature \label{base}}

\subsection{Oscillator Strengths}

This section describes the various literature sources that were considered for the base line 
list.  Criteria for accepting and rejecting various sources are presented.   
When we refer to the oscillator strengths, we use the standard expression 
log($gf$) for the base ten logarithm of the product of the lower level 
degeneracy and absorption oscillator strength.     

\subsubsection{Molecules \label{moles}}

The molecular line lists are taken almost entirely from literature sources. For DR10 an attempt was made to fit the molecular features 
to the very weak lines seen in the Sun in order to define solar $gf$ values. For the INT and DR12 line lists, 
the molecular astrophysical log($gf$) changes were removed in favor of 
adopting the best literature values available. Below we discuss separately each molecule and note references adopted, 
as well as other line lists that were tested.  

\setcounter{secnumdepth}{4}
\paragraph{CN}


We used the Kurucz (CD-ROM 18) CNAX.ASC and CNBX.ASC lists as the base line list. This combined
list was tested against the line list in \citet[MB99]{MB99}.   The latter is a hybrid of theoretical log($gf$) values
provided by S. P. Davis and astrophysical values fit to the Sun using abundances from \citet{Grevesse96}.
Our tests indicated that the line positions and strengths from MB99 were found to overall provide a better
match to the CN features in the Sun and Arcturus.   
Thus, the MB99 list was adopted, while we kept in our line list any of the CN lines which were in the Kurucz line list
and not in MB99.
We tested this composite list against the 
2010 version of the Plez CN line list (private communication) and found that our composite list better fit the Sun 
and Arcturus spectra, i.e., resulting in fewer poorly fit lines and fewer predicted lines without 
associated observable features.    For DR10 we
changed the CN molecules' log($gf$) values by +0.03 to fit the strongest lines in the Sun based on a visual inspection.
For INT and DR12 we removed the astrophysical log($gf$) offset and the composite line list (MB99 $+$ Kurucz) was adopted
in the final line list.

\paragraph{CO}

We tested the Kurucz (CD-ROM 18) COAX.ASC and COXX.ASC composite line list against the line list from 
\citet{Goorvitch} and found the latter to 
be slightly better in strength, but not in wavelength.  
Thus we adopt the Goorvitch log($gf$) values 
for the CO lines in the Kurucz list, while keeping the original wavelengths as in the Kurucz line list..
There were no lines in the Goorvitch list that were not in the Kurucz list.   We retained all lines in 
the Kurucz list that were not included in the Goorvitch list.  For DR10 we
changed the CO molecules log($gf$) values by -0.10 to fit the strongest lines in the Sun based on 
an overall visual inspection of the obtained fits.
For INT and DR12 we removed any astrophysical log($gf$) values and the Goorvitch and Kurucz hybrid line list was adopted.

\paragraph{OH}

We tested the Kurucz (CD-ROM 18) OH.ASC list  against that of \citet{Goldman} 
and found the Goldman lines to provide a better fit to the Sun and Arcturus.   
We adopted the OH lines from \citet{Goldman}  for our base line list.
For DR10 we changed the OH molecules log($gf$) values by -0.07 to fit the strongest lines in the Sun 
based on an overall visual inspection of the obtained fits.
For INT and DR12 we removed any astrophysical log($gf$) values, and the 
Goldman$+$Kurucz OH lines were adopted as the final product.

\paragraph{H$_2$}

The Kurucz (CD-ROM 18) H2.ASC was adopted as a base line list but was supplemented by a few additional lines from 
hdxx.asc (Kurucz web site, \\
\url{http://kurucz.harvard.edu/molecules/}).  
For DR12 we reduced the log($gf$) value for the H$_2$ line at 16586.664 \AA\ by 6.0 dex,  based on a poor fit to the Arcturus spectrum
and to be in line with the other log($gf$) values for the H$_2$ lines.

\paragraph{C$_2$}

The Kurucz (CD-ROM 18)  C2AX.ASC, C2BA.ASC, C2DA.ASC, and C2EA.ASC files were used as a base
line list.   For DR12 we fixed some C$_2$ AX features  that were clearly discrepant in Arcturus by 
replacing their log($gf$) values with those from \citet{Brault} and \citet{Kokkin}.

\paragraph{SiH}

We used the Kurucz (CD-ROM 18) HYDRIDES.ASC list for the SiH features as a base line list.  While these 
features should be visible in only very few giant and dwarf stars, they have been 
included in the line list for completeness.

\paragraph{\label{FeH} FeH}

For DR12, several FeH line lists were tested, including lines from \citet{Langhoff}  as implemented 
in the Uppsala spectral synthesis code BSYN, and \citet{Dulick} as implemented by Kurucz in fehfx.asc 
(Kurucz web site, \\
\url{http://kurucz.harvard.edu/molecules/}).   
We tested these on H-band NIRSPEC spectra of the cool dwarf stars GL436 \citep{Prato02} and GL763 
\citep{Bender05}, kindly provided by C. Bender.   Based on the synthetic spectra generated 
using these two line lists and stellar parameters, we expected to find a large number of weak
FeH features.   We searched for conspicuous spectral features in the observed spectra at locations
that both line lists predicted there to be relatively isolated FeH lines, but did not find 
any consistently accurate predictions.   In addition, we tried to determine if FeH lines would 
help in a statistical sense even if the strongest features could not be identified.  The first
test was a cross-correlation of a synthetic spectrum of pure FeH against the observed spectra; 
no peak was found at the rest velocity of either star.    The second test was to subtract the synthetic
spectra with and without the FeH features from the observed spectrum; in both cases the 
scatter increased with the inclusion of the FeH lines.   From these tests we conclude that 
the current FeH line lists would not assist in our analysis of the APOGEE spectra, thus 
the FeH lines were not included for DR12.

\subsubsection{Atoms \label{Atoms}}

For the atomic features, we compiled an atomic line list from a variety of different sources.   As a base line list, we started with 
the Kurucz line list gfhy3000.dat (Kurucz web site, \\
\url{http://kurucz.harvard.edu/linelists/gfhyper100/}).   
In this line list we included, in separate columns, lab data, 
the ``best'' empirical log($gf$) values in the literature,
and our astrophysical log($gf$) values.     This line list contains many more lines than are typically detectable in $H$-band stellar spectra
of cool giants, 
but all lines were retained in the line list as it may aid future investigations of extremely hot stars,
investigations of nebular features, or laboratory efforts.

\paragraph{Laboratory Data}

The gold standard for line list data is high quality laboratory measurements of energy levels, branching fractions and lifetimes.   
With this type of data we will have not only the best quality values but also quantifiable errors that can be used
to constrain empirical changes from laboratory values, as will be discussed in Section \ref{astro}.
In this section we consider both true laboratory measurements and any theoretical measurements that have 
such well constrained theoretical uncertainty as to allow uncertainties to be included.
For example, in the case of hydrogen (Paschen lines), the 
National Institute of Standards and Technology database \citep[NIST;][]{Kramida14} 
gives the uncertainties in the log($gf$) as AAA or  $\le	0.3\%$.
Within this paper we will refer to any log($gf$) that has quantifiable 
errors as laboratory data.  Table \ref{lab data} contains the origins of the different lab sources adopted in our list.
Most of these were compiled in the NIST database and we adopt their usual 
grade-to-uncertainty conversion (Table \ref{uncertainties}).
In the case of Ti I, we adopted log($gf$) values from \citet{LGWSC13} over 
\citet{BLNPJLPV06}  when available.  


\begin{table}
\caption{
\label{lab data}
Laboratory Sources for Oscillator Strengths}
\begin{tabular}{|l|c|}  \hline
Species & Source  \\
\hline
H I & NIST --  \citet{WF09} \\
He I & NIST --  \citet{WF09}  \\
C I & NIST -- \citet{WFD96}, NIST -- \citet{WF07} \\
C II & NIST --  \citet{WFD96}, NIST -- \citet{WF07}   \\
C III & NIST --  \citet{WFD96}  \\
C IV & NIST --  \citet{WFD96}  \\
N I & NIST --  \citet{WFD96}, NIST -- \citet{WF07}  \\
N II & NIST --  \citet{WFD96}, NIST -- \citet{WF07}   \\
N III & NIST -- \citet{WFD96} \\
N V & NIST --  \citet{WFD96}   \\
O I & NIST --  \citet{WFD96}  \\
O II & NIST --  \citet{WFD96}   \\
O III & NIST --  \citet{WFD96}   \\
Na I & NIST - \citet{KP08}, NIST -- \citet{S08}  \\
Mg I & NIST --  \citet{KP08} \\
Mg II & NIST - \citet{KP08}  \\
Al I & NIST -- \citet{KP08}\\
Si I & NIST --  \citet{KP08}  \\
Si II & NIST - \citet{KP08}  \\
Si III & NIST - \citet{KP08}  \\
S I & NIST -- \citet{PKW09} \\
Ar I & NIST - \citet{WSM69} \\
K I & NIST - \citet{S08}, NIST -- \citet{WSM69} \\
Ti I & \citet{LGWSC13}, NIST -- \citet{BLNPJLPV06}   \\
Ti II &  \citet{WLSC13} \\
V I & NIST -- \citet{S08}, NIST -- \citet{WSM69}  \\
Fe I &  \citet{RANP13} \\
\hline
\end{tabular}
\tablenotetext{}{The species with NIST indicated as  a Source contain the noted
references in the NIST database.}
\end{table}

\begin{table}
\caption{
\label{uncertainties}
Adopted Uncertainties for the Oscillator Strengths}
\begin{tabular}{|c|c|}  \hline
Grade & Uncertainty  \\
\hline
 AAA	& $\le	0.3\%$ \\
 AA	& $\le	1\%$ \\
 A+	& $\le	2\% $ \\
 A	& $\le	3\% $ \\
 B+	& $\le	7\% $ \\
 B	& $\le	10\% $\\
 C+	& $\le	18\% $\\
 C	& $\le	25\% $\\
 D+	& $\le	40\% $\\
 D	& $\le	50\% $\\
 E	& $>	50\%. $\\
\hline
\end{tabular}
\tablenotetext{}{Grade-to-uncertainty conversion for the oscillator strengths as defined by NIST}
\end{table}


In addition to the laboratory log($gf$) values from the sources above, 
for the DR12 line list we updated the theoretical wavelengths with 
wavelengths from the following sources:
\begin{itemize}
\item Ti I wavelengths from \citet{S12} 
\item Ti II wavelengths from \citet{S12}
\item V I wavelengths  from \citet{T11} 
\item Cr I wavelengths from \citet{S12}
\item Cr II wavelengths from \citet{S12}
\item Rb I wavelengths from \citet{S06} 
\end{itemize}

\paragraph{Literature Astrophysical and Theoretical  log($gf$) Values}

The goal of these efforts is for the line list to be a comprehensive list of all $H$-band transitions which may appear 
in APOGEE spectra.   With this 
goal in mind, we have taken the base line list (as defined in \ref{Atoms}) and augmented it with 
additional lines from various theoretical predictions.   Since we are not the first to generate $H$-band line lists,
we also tested some of these literature compilations against our base list to determine if we might improve
the list with  transitions from these sources.   
Since there are few $H$-band transitions known for elements heavier than copper, we 
placed extra emphasis and effort on $Z > 30$ transitions.       

A careful review of all pages of the Kurucz web site (\url{http://kurucz.harvard.edu/atoms.html})
revealed a few additional Ti, Fe, and Cr transitions, which were added to the base line list.
We tested the MB99,  \citet{R09}, and \citet{R10} line lists and found their log($gf$) values 
to provide a better fit to the Sun than the base line list.   
This should not be surprising as these three references are based on 
astrophysical solar log($gf$) values.  All of these log($gf$) values were added, giving preference to the most recent 
line lists, e.g., \citet{R09} over MB99, for any features in common between the lists.
We added Ce III lines from \citet{Wyart} and Y II lines from the DREAM compilation 
\citep[Feb. 2011 download]{Biemont}.  We replaced
all log($gf$) values for Ca I and Ca II 
with the values given in \citet{Hansen99},  and  \citet{Laughlin}, respectively.

\paragraph{Hyper-fine Splitting}

For DR12 we adopted the energy levels for V I lines from  \citet{T11}, and 
we added the V hyper-fine splitting (HFS) components using the coefficients of \citet{Palmeri97},
\citet{Palmeri95},   and \citet{Guzelcimen}.   
\citet{Guzelcimen14} and \citet{Wood14}  have published laboratory measurements for the V energy levels  since the 
DR12 line list was generated.    These values produce HFS components that are in excellent agreement with the earlier theoretical
predictions.
Based on the Na I components from (\citet{Safronova}, \citet{Happer74}) 
the HFS is too small to impact the line profile or abundance analysis.   

For Al the HFS can be significant and its exclusion from the APOGEE line lists deserves comment.
Using the HFS components from \citet{Falkenburg79} and \citet{Sur2005} we find that
for the weaker Al lines the impact of HFS is less than 0.1 dex (a typical uncertainty for APOGEE)
while for the strongest line at 16755 it has offsets as large as 0.4 dex for the coolest, most metal-rich
giants.    ASPCAP measures abundances from all of the Al features and thus will potentially have a bias toward
abundances which are slightly too high and a function of line strength (largely metallicity and effective
temperature).   Examination of the slope and scatter of the Al abundances for open clusters in the 
APOGEE calibration sample (Figure 5 in \citet{Holtzman2015} suggests that this bias is less than 0.1
dex.    A secondary check can be made by comparing literature determinations of Al in globular clusters
compared to those using an APOGEE line list as in \citet{Meszaros15}, Figure 4.  If HFS was a significant
influence in these stars then one would expect that the [Al/Fe] enhanced abundances would be biased positively
compared to the [Al/Fe] ''normal'' stars and that the affect would be strongest in the most metal-rich clusters.   
There is a small trend among the clusters of the order of 0.2 dex.  
While Al HFS will be added in future versions of the APOGEE line list,
users of this line list in the future should be aware of this limitation and perhaps
avoid the use of the strongest Al feature as was done in \citet{Smith13}.


\subsection{ Damping \label{damp}}

The line width for a spectral feature is a complicated function of stellar rotation, thermal broadening,
turbulence and surface convection,  and several other types of 
quantum mechanical broadening that mostly impact strong lines, including Stark, resonance, and van der Waals.    
Broadening coefficients are often included in line lists, and one of the most important of these
is the van der Waals broadening.
In this work, what we refer to as ``damping'' is actually the log of the 
broadening coefficient: van der Waals collisional damping divided by the 
number density of hydrogen, 
or  $log (\Gamma_6/N_H) = 17 v^{3/5} C^{2/5}_6 $,
where  $v$ is the velocity (set by thermal motions with T$_{eff} = 10000$)
and $C_6$ is the interaction constant.
There are only a few different sources for the damping or $C_6$ values in the $H$-band: 
the Kurucz compilation,  MB99, \citet{R09}, and \citet{R10}.
MB99 actually used several different methods to get the damping: fits to the solar spectrum,  
values out of \citet{BAO98} and other references.

The Barklem web site \citep[\url{http://www.astro.uu.se/$\sim$barklem/}, v2.0 ][]{BAO98} has 
codes for calculating the van der Waals damping. 
However, these codes only work for certain values of the effective principal quantum number. 
We downloaded v2 and ran it for all of the transitions within range.
For many of the IR transitions the effective principal quantum number is outside of the Barklem grid.  

It is possible to approximate the $C_6$ values as $6.46E-34 \Delta r^2$,
where   $\Delta r$ is the unit-less difference in the mean square radius of the two energy levels \citep{Unsold}. 
The radius can be described roughly as $r = 0.5 * n^2 * (5* n^2 + 1 - 3*l(l+1))$ (MB99)
where $n = 1 / ( \Delta EP / \chi$)$^{0.5}$,
$\Delta EP$ is the difference between the energy level of the transition and the ionization energy ($\chi$), and 
$l$ is the orbital angular momentum quantum number.   This approximation 
yields a $C_6$ value 10 times larger than that calculated by \citet{Unsold}.

\begin{figure}
\includegraphics[scale=0.9,origin=c]{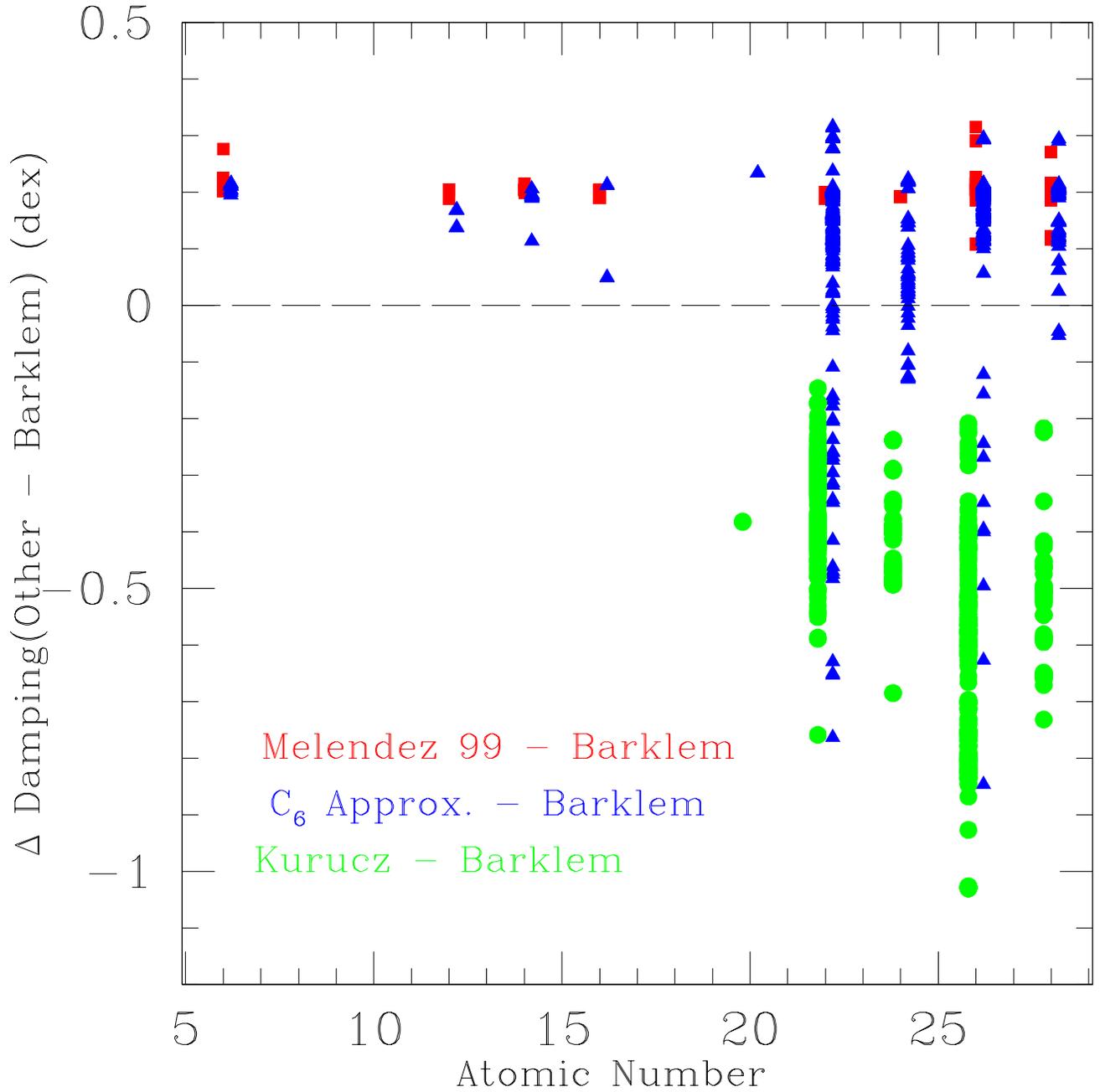}
\caption{
\label{Dampingcomparison}
Difference between the log damping values from various sources and the Barklem values. 
The x axis shows atomic number. 
The red squares are MB99 - Barklem, the blue triangles are $C_6$ 
approximation - Barklem, and the green circles are Kurucz - Barklem.  The point type 
has been offset slightly to avoid them overlapping each other. 
The dashed line at  $\Delta = 0$ is provided to guide the eye.
}
\end{figure}

We assumed the Barklem values to be the best possible source for theoretically derived damping constants.   In Section \ref{astrodamp} we 
show that these theoretical values are very close to the calculated values based on comparisons to the solar spectrum.  
Figure \ref{Dampingcomparison} shows the difference between the different damping constants and the Barklem values as a function 
of atomic mass number for those lines in common.   The Kurucz damping constants are systematically too small 
by several tenths of a dex, while the $C_6$ approximation and MB99 values are too large by a 
few tenths of a dex.   From this assessment, our ranked preferences for adopted damping values for the base line list are:
\begin{enumerate}
\item \citet{BAO98} whenever possible.
\item \citet{R09} and \citet{R10}.
\item MB99.
\item The $C_6$ approximation given in this section.
\item Kurucz values
\end{enumerate}

\section{Semi-automated ``Astrophysical'' Line List Calculations \label{astro}}

It is possible to improve upon theoretical log($gf$) and damping values by comparing line profiles and strengths of 
well understood spectra to synthetic spectra.    Even laboratory measurements with significant error bars could be 
improved by this type of comparison if done within the errors of the measurement.   We refer to these corrections based
on observed spectra as ``astrophysical'' line parameters.   In this section we describe the software and methodology 
we created to generate astrophysical line parameters and compare those parameters to the base line list values.  
It should be noted that in making changes to the atomic line parameters to match the observations one is also masking 
systematic errors in the model atmospheres and the line formation calculations.   

\subsection{Astrophysical Software and Methodology}

We have developed a code that can vary the log($gf$) and damping values in a line list
to match one or more observed spectra.   The final product of the code is a set of astrophysical log($gf$) 
and damping values that we adopt in the final line list.   
This code relies on having accurate stellar parameters for the observed
stellar spectra and a constrained and accurate base line list.      
The code was developed within the APOGEE team by DB and is available through the 
Astrophysics Source Code Library \citep[][]{2015ascl.soft02022B}.
This code has
evolved, and the version used for DR12 is more complex than that used for DR10.   Below 
we document the state of the code as implemented for DR12 and describe the significant changes 
that were made after DR10.   

The code is written in IDL as a wrapper around the spectral synthesis code MOOG \citep{MOOG}. 
For DR10 the 2009 version of MOOG was used, and for DR12 the 2013 version was used.  \citet{Zamora15} 
conducted an analysis of the differences between synthetic spectra computed with both MOOG and ASS{$\epsilon$}T and 
found the differences to be very
minor for the Sun and Arcturus, $< 0.01$ except around the hydrogen lines.
The astrophysical $gf$-fitting code starts with an input line list, which in our case is the best laboratory and literature
values available between 1.5 and 1.7 $\mu m$ (see Section \ref{base}).
For spectral comparison we adopt an $H$-band center-of-disk (COD) atlas for the Sun 
\citep{LW91} and an Arcturus flux atlas \citep{Hinkle95}.   
The adopted solar and Arcturus stellar parameters and abundances are 
listed in Tables \ref{models} and \ref{modelabund}.
The DR10 solar model was created from a grid of Kurucz Atlas models \citep{Castelli97} with 
computed Opacity Distribution Functions (ODFs).  
We interpolate within this grid using the code ``makekuruczpublic'' \citep{McWilliam95}.
Unfortunately, after the DR10 calculations were made, a mistake entered into our methodology and the 
synthesis of the Sun was done as if it were a full flux spectrum, rather than a 
COD calculation.   This is 
noted in Table \ref{models} as ``not COD''.   In Section \ref{COD} we discuss the 
impact of this on the INT
and DR12 line lists and how it may impact the subsequently derived abundances.  

The DR10 
Arcturus model atmosphere was interpolated from the 2005 MARCS grid
\citep[][, further expanded by B. Edvardsson 2015, private communication]{Gustafsson03, Gustafsson08}.    
The MARCS models have opacity sampling instead of ODFs; however, the MARCS models are spherical, 
with appropriate $\alpha$-enhancement at the metallicity of Arcturus, and thus were deemed to 
be more appropriate for DR10.
Although the differences between the most recent Kurucz and
MARCS models are minimal at the effective temperature and gravity of Arcturus \citep{Zamora15},
we adopted a Kurucz model for
Arcturus after DR10 to be consistent with the rest of the ASPCAP models.     
For the INT line list,
we use the same Kurucz grid from which we pulled the solar model.   We should note that the 
exact details of the Arcturus model atmosphere for DR10 and INT will have little impact on the 
final line list because, as will be described in more detail below, the final solar log($gf$) fitting removed
all but the weakest or most temperature/gravity sensitive features in the Arcturus log($gf$) fitting.   For DR12
the log($gf$) fitting methodology changed so the model was far more critical.   For DR12 we adopted a 
model generated in the same way as the model atmosphere grid points were generated for ASPCAP, namely
from the Kurucz Atlas 9 code \citep[see][]{Meszaros12}.

\begin{table}
\caption{
\label{models}
Model Atmospheres Adopted for Astrophysical log($gf$) Value Calculations}
\begin{tabular}{|r|r|r|r|r|r|}  \hline \hline
Model & $T_{eff}$ & log(g) & [Fe/H] & $v_t$ & Notes \\
\hline
\multicolumn{6}{c}{Sun}\\
\hline
DR10 & 5780 & 4.40 & +0.00 & 1.10 & Kurucz  ODFNEW \\
INT    & 5777 & 4.44 & +0.00 & 1.10 & Kurucz  ODFNEW; not COD\\ 
DR12  & 5777 & 4.44 & +0.00 & 1.10 & Kurucz ODFNEW; not COD \\
\hline
\multicolumn{6}{c}{Arcturus} \\
\hline
DR10  & 4290 & 1.55 & -0.55 & 1.67 & MARCS \\
INT    & 4286 & 1.66 & -0.52 & 1.70 & Kurucz NOVER  \\
DR12  & 4286 & 1.66 & -0.52 & 1.70 & new revised Kurucz  \\
\hline
\end{tabular}
\tablecomments{
Not COD: these syntheses were mistakenly
calculated as full flux calculations within MOOG and not set as center of disk (COD).}
\end{table}

The solar abundances for DR10 and INT were adopted from \citet{Asplund09},  while for 
DR12 we adopted the abundances from \citet{Asplund05}.  This change was made to 
make the assumed abundances in the line list consistent with those adopted in the Kurucz model atmospheres 
calculated for APOGEE \citep{Zamora15, Meszaros12}.  

The Arcturus abundances were set with a number of considerations.   
Since we were concerned about 
misfitting the atomic features near molecular features, 
we forced the C, N, and O abundances
to match those of \citet{Smith13}, who derived those 
abundances in an independent analysis in the $H$-band using line list INT.
The C, N, and O abundances from \cite{Smith13} are very similar (within $\sim$0.05 -- 0.10 dex)
to those derived recently by \citet{Abia2012}, who used NIR lines of CO, OH, and CN, as
well as \citet{Sneden14}, who used optical C$_{2}$, [O I], and CN transitions.  The
mean and standard deviations of these 3 studies are A(C)=8.01$\pm$0.05, A(N)=7.66$\pm$0.02,
and A(O)=8.64$\pm$0.03, with $^{12}$C/$^{13}$C=7.4$\pm$1.4; the Arcturus C, N, and O abundances
are therefore well-constrained in these recent independent studies.  
We wanted the abundances to be largely self-consistent with those adopted in the 
model atmosphere, so the abundances
of Mg, Si, S, and Ti were all given $\alpha$-enhanced values of $+0.4$ above scaled solar, 
whereas all of the odd Z elements (except Al) and iron peak elements were
given scaled solar values.  Al has several
very strong features with several other important features in its wings, thus we adopted a value to 
fit all three strong lines.  This adopted value was close to that of \citet{Smith13}, who
excluded the strongest line from their analysis. 
The abundance of Ca clearly deviates significantly from the other alpha elements,
so we adopted a value close to that of \citet{Smith13}. 
For DR10 the Arcturus log($gf$) values were only adopted 
for a small subset of those lines-of-interest (LOI) 
which were above the threshold in Arcturus but not in the Sun.
For DR12 the adopted Arcturus abundances play a far more significant role because the final Arcturus and solar log($gf$) values
were averaged together, although the solar values were given more weight.  
In retrospect, the self-consistency requirement, i.e., forcing the abundances to match the model, should 
not have been a significant driver and may have resulted in a less optimal line list that resulted in an abundance 
inconsistency for Arcturus in ASPCAP \citep[see Section 5.5.1 in ][]{Holtzman2015}.

\begin{table}
\caption{
\label{modelabund}
Abundances Adopted for Astrophysical log($gf$) Value Calculation}
\begin{tabular}{|r|r|r|r|r|r|r|r|}  \hline \hline
\multicolumn{2}{c}{Atomic} & \multicolumn{3}{c}{Sun} & \multicolumn{3}{c}{Arcturus }\\
Number&Species&DR10 & INT & DR12 & DR10 & INT & DR12 \\
\hline
6   & C   & 8.43 & 8.43 & 8.39 & 7.88 & 8.27 & 7.96 \\
7   &  N & 7.83 & 7.83  & 7.78 & 7.21 & 7.26 & 7.64 \\
8   &  O  & 8.69 & 8.69  & 8.66 & 7.59 & 8.54 & 8.64 \\
11 & Na   & 6.27 & 6.27  & 6.17 & 5.73 & 5.65 & 5.65 \\
12 & Mg   & 7.53 & 7.53  & 7.53 & 7.24 &  7.41 & 7.41 \\
13 & Al   & 6.43 & 6.43  & 6.37 & 6.24  & 6.15 & 6.15 \\
14 & Si   & 7.51 & 7.51  & 7.51 & 7.02 & 7.39 & 7.39 \\
16 & S   & 7.15 & 7.15  & 7.14 & 6.70 & 7.02 & 7.02 \\
19 & K   & 5.08 & 5.08  & 5.08 & 4.82 & 4.56 & 4.56 \\
20 & Ca   & 6.29 & 6.29  & 6.31 & 5.85 & 5.89 & 5.89 \\
22 & Ti   & 4.91 & 4.91  & 4.90 & 4.69 & 4.78 & 4.78 \\
23 & V   & 3.96 & 3.96  & 4.00 & 3.41 & 3.48 & 3.48 \\
24 & Cr   & 5.64 & 5.64  & 5.64 & 5.09 & 5.12 & 5.12 \\
25 & Mn   & 5.48 & 5.48  & 5.39 & 4.69 & 4.87 & 4.87 \\
26 &  Fe  & 7.45 & 7.45  & 7.45 & 6.90 & 6.93 & 6.93 \\
27 & Co   & 4.87 & 4.87  & 4.92 & 4.32 & 4.40 & 4.40 \\
28 &  Ni  & 6.20 & 6.20  & 6.23 & 5.65 & 5.71 & 5.71 \\
\hline
\end{tabular}
\end{table}

In order to match the line broadening in the Arcturus atlas spectrum, we
convolved all synthetic spectra with an instrumental profile and 
a rotational profile ($v$sin$i$ = 2.0 km s$^{-1}$) \citep{Gray1981,Gray2006}. The limb darkening
for Arcturus in the NIR is assumed to be 0.46 \citep{claret00}. Solar synthetic
spectra were compared with the COD atlas spectrum, and
therefore were not corrected for rotation but they were convolved with an
instrumental profile.   The instrument profile 
was determined from an analysis using cross-correlation against a synthetic spectrum
generated from the base line list but avoiding the strongest lines.

\begin{table} 
\caption{
\label{line_depth}
Number of strong lines in the solar and Arcturus spectra.}
\begin{tabular}{lrrrr}
\tableline\tableline
 Limiting &  \multicolumn{2}{c}{Sun} & \multicolumn{2}{c}{Arcturus} \\
 depth    &  all lines    & atomic only  &          all lines  & atomic only \\
\tableline
1E-4  & 10775 & 3409  & 31192 &  4287 \\
5E-4  &   5600 & 2426  & 19731 &  2819 \\
1E-3  &   4291 & 2166  & 15919 &  2559 \\
1E-2  &   1348 &   989  &   8758 &  1590 \\
 \tableline   
 \end{tabular}
 \tablecomments{
The number of strong lines (atomic and molecular, or atomic only) in the synthetic 
spectra of the Sun and Arcturus in the range between 1.50 and 1.70 $\mu$m in 
dependence of the limiting line depth.
}
\end{table}

For DR10 and INT we selected the LOI 
using the line strength (opacity at line center) tabulation
built into MOOG which allowed us to remove extremely weak lines.    
For DR12 we refined this by running each line individually and noting
the line depth.    The number of strong lines in a spectrum varies significantly with stellar
spectral type and with an adopted cutoff for the minimum line strength. 
We calculated the maximum line depth with respect to the normalized continuum
for each line in the initial line list. Table \ref{line_depth} shows the number 
of atomic and molecular features deeper than the listed depth in the solar
and Arcturus spectra. Since the target S/N ratio for APOGEE program stars 
is 100, 
we considered evaluating only those lines that were deeper than 0.001 with respect to the 
normalized continuum.   For reference the weakest lines visible in Figure \ref{show_fit} 
have depth of 0.004 and equivalent widths of $<$0.6m\AA\  and are well below the detection threshold 
in APOGEE spectra.   The final set of LOI is the joint list 
for the Sun and Arcturus corresponding to the $>$0.001 level (see Table \ref{line_depth}).
Future improvements to the line fitting code will likely involve the entire line strength
and not the depth at the line center as this would remove line depth variations
due to damping and HFS differences from line to line.

Once we had the LOI, we started evaluating them one
by one in order of line strength, with the strongest line first. 
For each line we fit the spectral range within 0.8 \AA\ around the line center. To account for 
possible wings of nearby strong lines, we calculated the contribution of all lines within 
18 \AA\ around each considered LOI. We evaluated two free parameters: radial velocity
changes within $\pm$ 0.25 km s$^{-1}$ (which also accounts for the uncertainty in the central wavelength)
and log($gf$), which is allowed to vary within the following set of rules:
\begin{enumerate}
\item for DR10 and INT, the lines with measured laboratory values were allowed to vary within 1 sigma  of 
those laboratory measurements, and all other LOI were allowed to vary by -2 to +0.75 dex, and
\item for DR12, the lines with measured laboratory values were allowed to vary within 2 sigma  of 
those laboratory measurements, and all other LOI were allowed to vary by -2 to +0.75 dex.
\end{enumerate}
The reasoning behind the asymmetric limits on the log($gf$) variation is motivated by the fact that
there are bad and missing lines in the base line list.
There are a number of strong theoretical lines without observed counterparts, and we want these lines
to be strongly suppressed, which sets the lower limit of $-2$ dex in log($gf$).   There are also a number of observed lines 
without theoretical or laboratory counterparts in the base line list, and we do not want to inflate inappropriate very 
weak lines, which sets the upper limit of $+0.75$ dex in the log($gf$).
The line list was updated with new log($gf$) values on the fly during the fitting.  
Spectral pieces around each LOI were
fitted using the downhill simplex optimization algorithm by \citet{NM65}.  Because 
the LOIs often overlapped, we performed the evaluation iteratively. 

The treatment of HFS changed between the INT line list and the DR12 line list.
Before DR12, no consideration was made to force the different components of an adjusted line to scale
with each other.    Thus, if the code required the strongest component of a line with HFS 
to be adjusted, then no change was made for weaker components.  
For DR12, we changed HFS component log($gf$)
values by the same amount. 

For DR10 and INT the majority of the LOI were fit only to the Sun,
but those lines that were not visible in the Sun (i.e., were below the depth threshold)
but were visible in Arcturus were fit to Arcturus.   LOI that were above the 
detection threshold in the Sun were fit again and superseded any corrections that were
based on the Arcturus spectrum.  Thus, LOI were either
fit to Arcturus or the Sun.  Atomic lines weaker than the cutoff, i.e., not a LOI, in both the 
Sun and Arcturus were left unchanged from the input line list's values.
So for DR10 and INT, the following methodology was adopted:
\begin{enumerate}
\item Fit the LOI in the Sun for two iterations for the log($gf$) values.\\
\item Fit the LOI in Arcturus for a single iteration for the log($gf$) values.\\
\item Fit the LOI in the Sun for two iterations for the log($gf$) values.\\
\item Fit the strongest LOI in the Sun for damping parameters.\\
\item Fit the LOI in the Sun for three final iterations of the log($gf$) values.\\
\end{enumerate}

For DR12 we implemented the use of both the solar and Arcturus spectra
to constrain the astrophysical log($gf$) values.  To accomplish this we determined the 
astrophysical values from Arcturus and from the Sun independently instead of 
serially (as was done for DR10 and INT) 
and then weighted the solutions, 
with the Sun getting twice the weight as Arcturus because the abundances in the solar 
photosphere are also confirmed by the meteoritic values.
Other weighting schemes are possible and will be considered for future line lists.
 


We noticed that most values of log($gf$) settled down quickly after the second iteration
over all LOI.  We also evaluated the damping constant for
the strongest lines, applying the same algorithm as for the derivation of the log($gf$) values 
but using only the solar spectrum
for the comparison. Our standard sequence of the line parameter adjustment 
process was 2 x $E$log($gf$), 2  x $E$damp, n x $E$log($gf$), where $E$log($gf$) and $E$damp designate 
one iteration over the whole LOI list for log($gf$) and for the damping constants, respectively. 

\begin{figure}
\epsscale{1.0}
\plotone{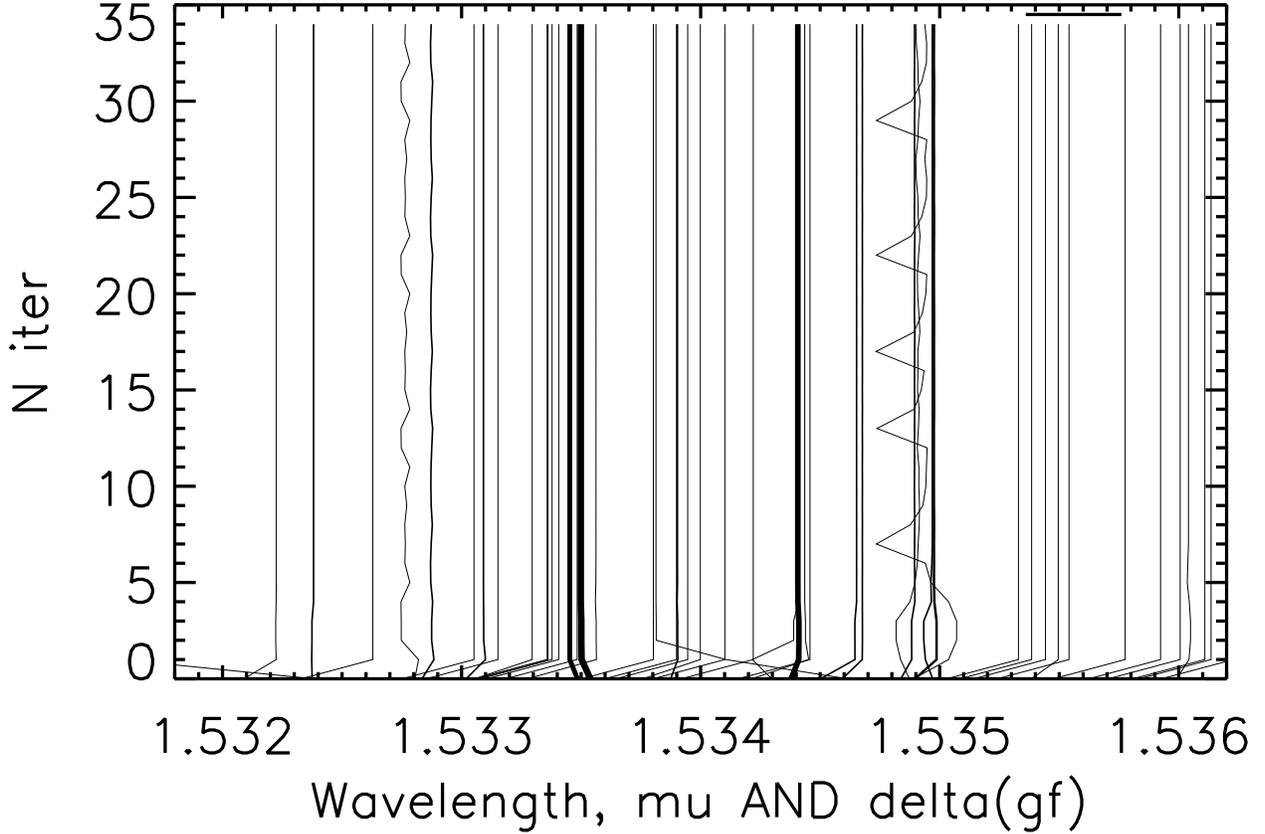}
\caption{
\label{hair_diagram}
Example of log($gf$) stability during fits to the Sun and Arcturus. 
The X-axis for this diagram is in microns combined with a scaling of offsets in log($gf$).  
The horizontal bar at the top right corner corresponds to the length of a 1.0 dex change in log($gf$). 
Change of the log($gf$) with respect to the original value is shown as the deviation from the vertical
line. The Y-axis shows the progress of the iterations: 2 x $E$log($gf$), then 2  x $E$damp, and then $n$ x $E$log($gf$).
Parameters of most of the lines settle down after the 3rd log($gf$)-iteration. The line thickness denote
the line depth with the thinest lines having depths less than 0.01 and the thickest lines having depths
greater than 0.2.
}
\end{figure}

\begin{figure}
\epsscale{0.6}
\plotone{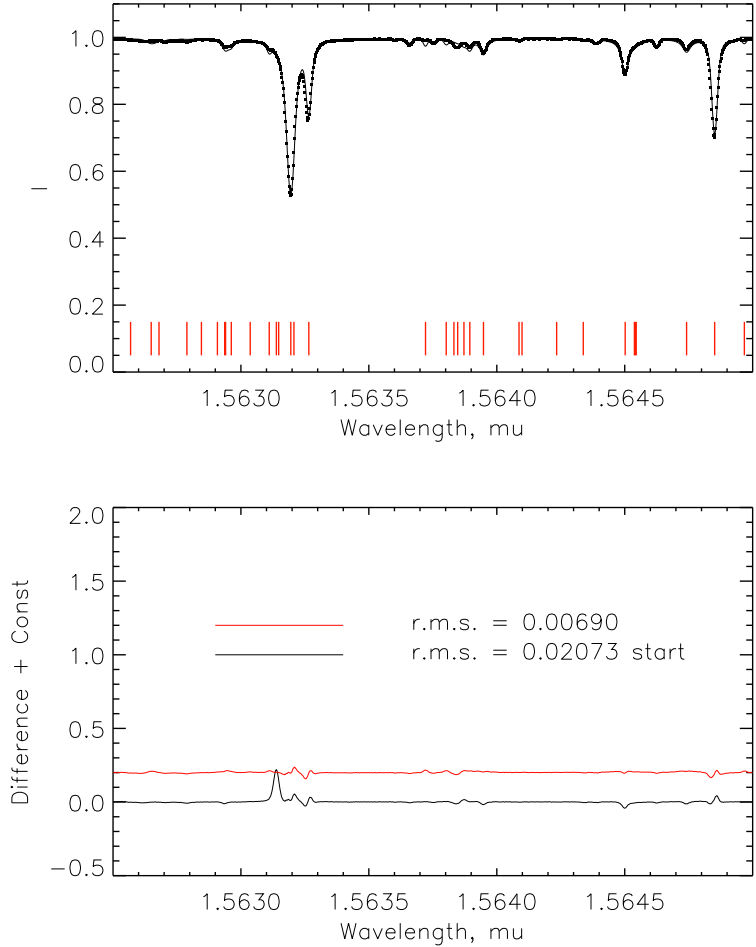}
\caption{
Example of the input and output spectra from the astrophysical adjustment of the line parameters.
Top panel: observed solar spectrum (dots) and best-fit synthesized 
spectrum (solid curve). The vertical red bars designate wavelengths of lines whose parameters
were adjusted.  Some visible features are not fit because they are molecular features, and for the
INT and DR12 line lists we made no astrophysical corrections to molecules.
Bottom panel: difference between the observed and synthetic solar spectra before (lower curve)
and after (upper curve) the iterations. The upper curve has been  shifted by $+0.2$ along the vertical axis. 
The weakest features visible in the observed spectrum have depth of 0.004 from the continuum and equivalent
widths of $<$0.6m\AA, at the resolution and SNR of APOGEE spectra these features would not be visible.
\label{show_fit}
}
\end{figure}

We justified the number of necessary iterations by running a few long sequences
``2 x $E$log($gf$), 2  x $E$damp, $n$ x $E$log($gf$)'' with n over 30. It was noticed that the log($gf$) values
for most of the LOIs settled down after the ``2 x $E$log($gf$), 2  x $E$damp, 2 x $E$log($gf$)'' sequence.
However, the log($gf$) in a few dozen of the LOIs did not settle down even after 30 iterations.
Instead, the parameters oscillated around certain values. This usually
happened to overlapping lines with similar depth.  We identified these lines in the last 
iterations in the ``2 x $E$log($gf$), 2  x $E$damp, 4 x $E$log($gf$)'' sequence and replaced 
these log($gf$) values with their average values. Figure \ref{hair_diagram} shows a ``hair-diagram''
for a small piece of spectrum with an illustration of log($gf$)-stable and unstable lines. Each curve
starts at its central wavelength at the x-axis. The x-deviation from the vertical designates the difference 
of log($gf$) from the original value. The vertical direction shows the iterative sequence progress
(2 x $E$log($gf$), 2  x $E$damp, and then $n$ x $E$log($gf$)). 

The version of MOOG we employed to generate the astrophysical line parameters 
does not have a proper treatment for the damping of  hydrogen lines.
Thus these features were artificially 
removed from the atlases by dividing by a synthetic spectrum containing only hydrogen lines, 
where these broad features were forced to fit.   The 
hydrogen lines were also removed from the input line list when importing it into MOOG;  the
final adopted line list used by APOGEE included the hydrogen lines.   Molecular lines could be adjusted
in a way similar to that adopted for the lines with HFS, where lines from the same multiplet or 
all multiplets could be adjusted by the same amount.   
We tried fitting the extremely weak CO, CN, and OH features in the Sun by eye for DR10.  
Given the weakness of the lines in the Sun,
no attempt to modify the literature
molecular log($gf$) values were made for INT and DR12, so
the final adopted line list is a result of an evaluation of only the atomic lines, 
excluding hydrogen.  Figure \ref{show_fit} shows an example of the fitted solar spectrum.   The rms scatter 
in the difference spectra (lower panel) is reduced from 0.021 with the base line list to 0.007 
with the final DR12 line list.   This improvement is seen across the spectrum and is a strong indication
that this methodology works.   
The most important test for any line list is to verify that it produces reliable abundances results requiring no or
little zero point corrections when compared with well established abundance trends and values from
the literature. We refer to  \citet{Holtzman2015} for such a discussion.

\subsection{Astrophysical log($gf$) Values}

\begin{figure}
\epsscale{0.8}
\includegraphics[scale=.50,angle=-90]{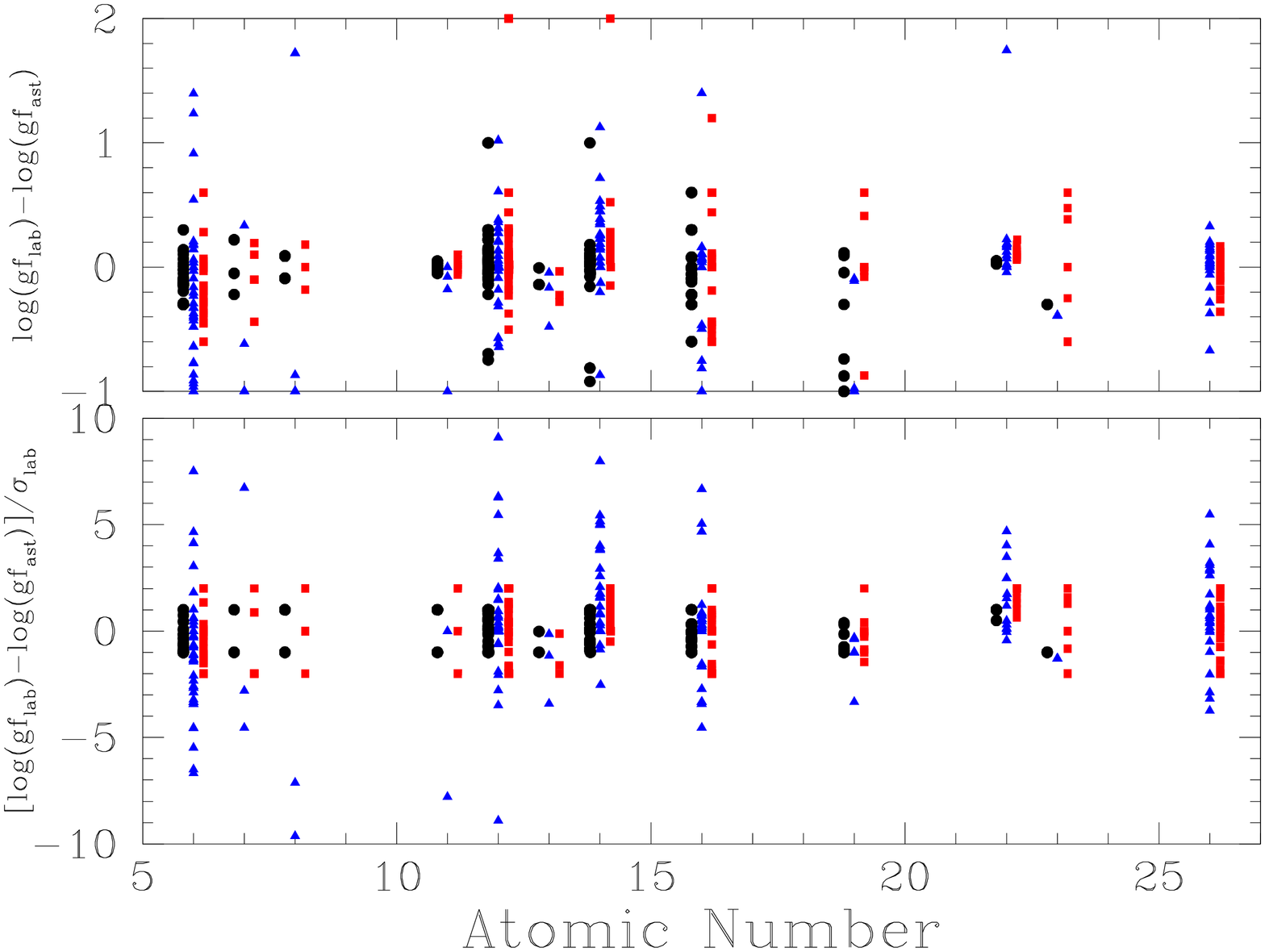}
\caption{
\label{labgfdiff2}
A comparison of the astrophysical and laboratory log($gf$) values from DR10 and DR12.  The top panel shows the differences between 
the log($gf$) values for each line and the bottom panel shows the difference between the log($gf$) values compared to the laboratory error in
the log($gf$) value. 
In this figure the black circles represent the atomic lines with lab data from DR10 and have been offset by -0.2 in atomic number for better visibility, 
the red squares represent the lines with lab data from DR12 and have been offset by $+0.2$ in atomic number for better visibility, 
and the blue triangles were calculated without any lab constraints using both Arcturus and the Sun with limits of -2 and +0.75 dex in log($gf$).
}
\end{figure}

As described above, the astrophysical log($gf$) values were calculated from lab log($gf$) values (when available), taking into consideration 
the errors in those measurements.
For DR10 and INT the astrophysical log($gf$) values were allowed to vary within one sigma of the 
laboratory errors, while for DR12 this was 
expanded to two sigma.   Figure \ref{labgfdiff2} shows the difference between the astrophysical and laboratory log($gf$) values (top panel) 
and this same difference divided by the laboratory error (bottom panel).    
For reference, the blue triangles in this figure show how much the log($gf$) values would change if we adopted
the same criterium used for the literature empirical log($gf$) values,  i.e., allowed changes of -2 to +0.75 dex.  
While some of the red squares and black circles in the lower panel of 
Figure \ref{labgfdiff2} do fall at the one and two sigma limits, 
most are within the error limits and centered around zero, which lends
confidence that these astrophysical log($gf$) values are of good accuracy.   The exception may be Ti, where the laboratory log($gf$) values tend to be larger than
the astrophysical log($gf$) values in both DR10 and DR12.    This may be caused by the temperature sensitivity of the Ti I lines and thus
were driven by the assumed abundance of Ti in Arcturus (the cooler of the two stars constraining the astrophysical log($gf$) values).  The actual 
errors on the laboratory Ti I log($gf$) values are very small and therefore the astrophysical log($gf$) values were not allowed to vary substantially
from the original laboratory log($gf$) values.    When the laboratory constraints were removed for Ti I
(the blue triangles in the upper
panel of Figure \ref{labgfdiff2}), 
the resulting log($gf$) values were again not  substantially different from the laboratory values, with the single
exception of  the Ti I transition at 16413.029 \AA.    

\subsubsection{Impact of using COD vs. Total Flux \label{COD} }

\begin{figure}
\includegraphics[scale=.50,angle=-90]{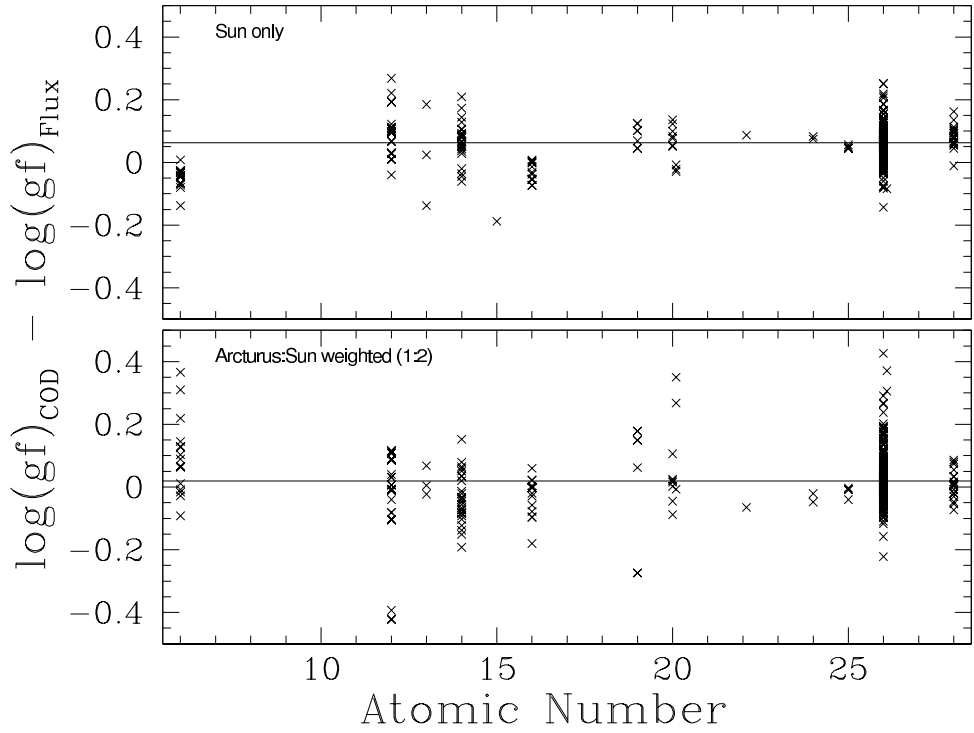}
\caption{
\label{CODdiff1}
The difference between the astrophysical log($gf$) values 
calculated with the proper center-of-disk (COD) option set in MOOG and the ``normal'' flux option set in MOOG, shown as a 
function of atomic number.   The top panel shows
the differences for a fit only to the Sun, as was done for INT, and the bottom panel shows the differences for a fit using 
both the Sun and Arcturus, as was done for DR12.  The solid line in
both panels is the best fit to the data.
}
\end{figure}

\begin{figure}
\includegraphics[scale=.50,angle=-90]{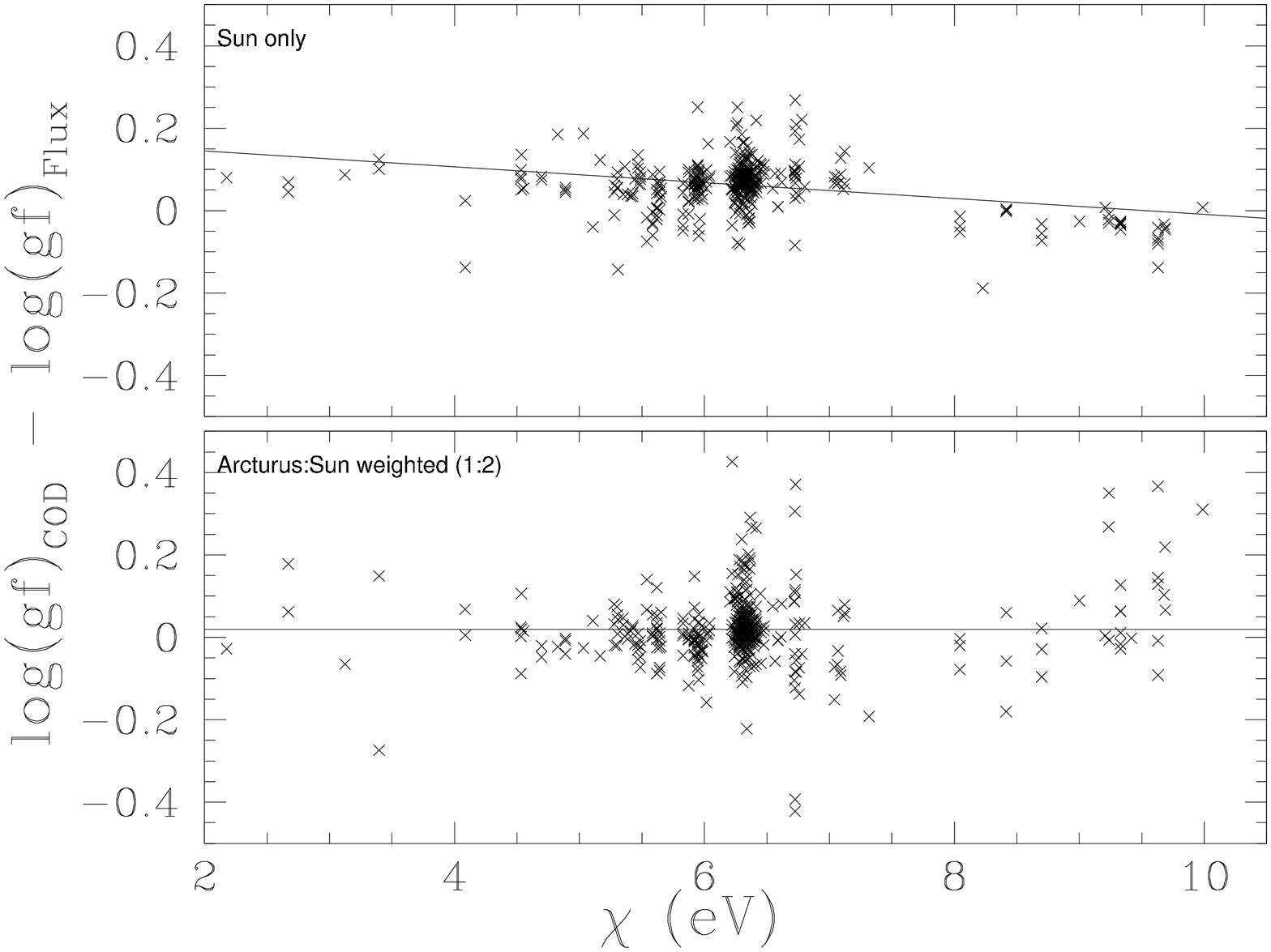}
\caption{
\label{CODdiff2}
The difference between the astrophysical log($gf$) values 
calculated with the proper center-of-disk (COD) option set in MOOG and the ``normal'' flux option set in MOOG, shown as a function 
of the lower level excitation potential.   The top panel shows
the difference for a fit only to the Sun, as was done for INT, and the bottom panel shows the differences for a fit using 
both the Sun and Arcturus, as was done for DR12.  The solid line in both panels is the best
fit to the data.
}
\end{figure}

As mentioned earlier, due to an incorrect entry in the MOOG parameter file, that was recognized later on, the solar log($gf$) values
of the INT and DR12 line lists were calculated in total flux rather than as COD, even though the reference solar spectral atlas is COD.    
Figures \ref{CODdiff1} and \ref{CODdiff2} show the impact of the difference in the astrophysical log($gf$) values 
calculated with the proper COD option set in MOOG's parameter file and the ``normal'' flux option. 
The top panel shows
the log($gf$) differences between COD and total flux 
for a fit only to the Sun, as was done for INT, and the bottom panel shows the differences for a fit using 
both the Sun and Arcturus, as was done for DR12. 
The Arcturus residuals are best fit with a constant offset of 0.019 dex in both figures.  
The INT residuals are best fit with a second order polynomial with respect to the 
excitation potential or a constant offset of 0.063 dex as a function of atomic numbers.
This shows that by not using the appropriate COD flag set in MOOG's parameter file is more problematic for the 
INT line list than for DR12.   The Sun-only fit could introduce a small temperature 
bias; a simple test using MOOG and these two line lists suggests a temperature offset of $\sim 40 K$ 
and a modest metallicity bias of $\sim -0.05$ dex, in the
sense that the log($gf$) values should have been larger and thus the derived abundances are too large.
\citet{GarciaPerez13} adopted the INT line list and thus was impacted by both of 
these zero points while
\citet{Smith13} and \citet{Cunha2015} are impacted by only the metallicity bias 
because their stellar effective temperatures and surface gravities were
 set with an independent methodology.   The exact impact these biases have are somewhat dependent upon the lines adopted.  
For example,
 all abundances based on molecules should not be influenced because these lines were not adjusted after DR10 
 in the astrophysical log($gf$) 
 methodology.    Lines with laboratory constraints may not have been impacted depending 
on whether the astrophysical log($gf$) derived was 
 at the limits imposed by the laboratory errors.    
 
 In summary, the mean impact of this improper treatment on DR12 is less than 0.02 dex on
the log($gf$) values.  In addition, there is no apparent slope with excitation 
potential or atomic number.  Since ASPCAP determines stellar parameters using
chi squared fits to the entire spectrum, the log($gf$) values for the molecular
and hydrogen lines are unchanged by the astrophysical log($gf$) corrections,
and the errors introduced by using the solar spectrum are, on
average, very small, we conclude that the impact is well within the errors 
presented in DR12.  

\subsection{Astrophysical Damping Values \label{astrodamp}}

\begin{figure}
\epsscale{0.8}
\plotone{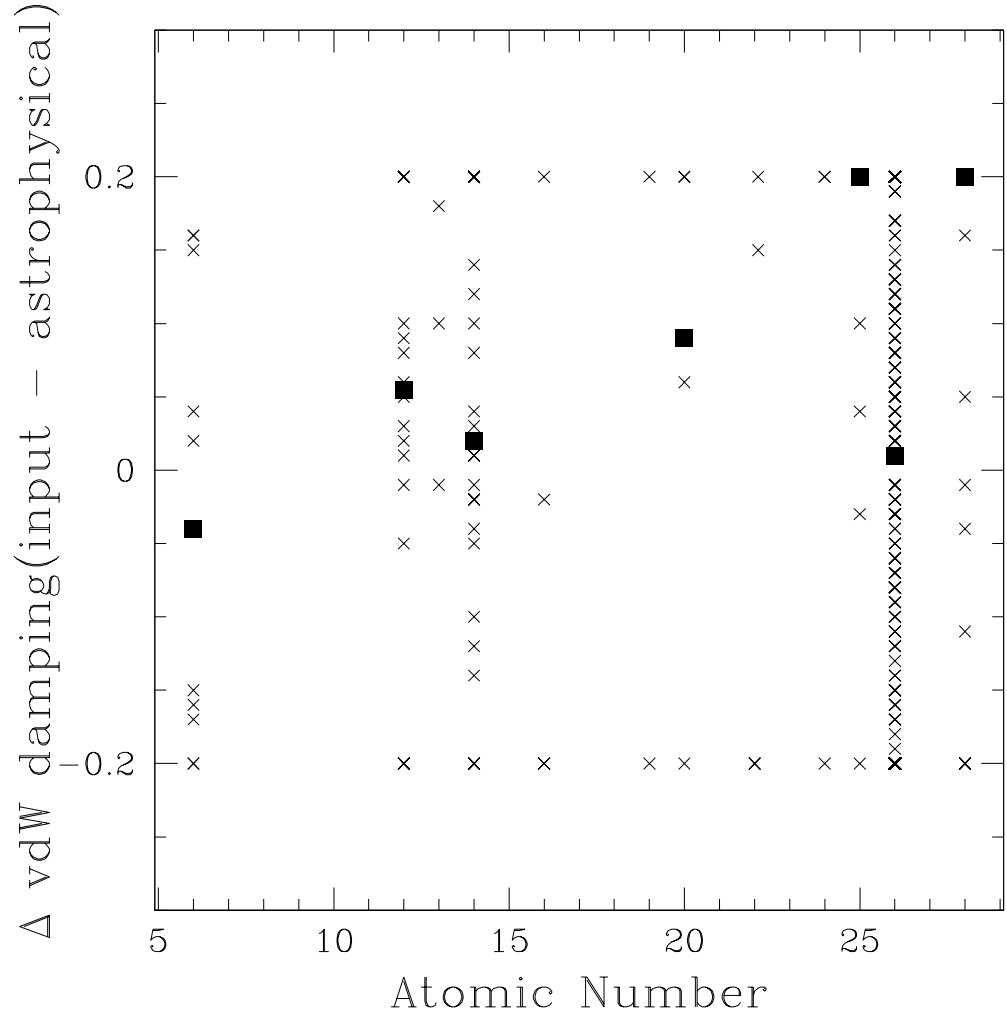}
\caption{
\label{DR12damping}
The difference between the input literature damping and the  astrophysical damping values 
 for DR12.  For elements with five or more points the median value is shown as a solid square.
}
\end{figure}

As mentioned in the previous section, after several iterations of the astrophysical log($gf$) code we fit the line profile for the 
strongest lines by allowing the
lines to be adjusted up to $\pm$ 0.2 dex from the input damping value.  Figure \ref{DR12damping} shows the difference 
between the input and astrophysical damping values.   The average difference for all damping values is 0.004 dex.  
Thus, on average, the 
astrophysical code is not making large changes.  However, it makes a small systematic change
to a few elements; for example, among Mg lines the average difference is 0.04 dex (the effect of damping is enhanced), 
and for the Fe lines 
the difference is $-0.04$ dex.

\subsection{Performance of the Astrophysical Parameters}

Over the full range of the line list the use of astrophysical 
log($gf$) values improves the fit in both the Sun and Arcturus.  
For the DR12 line list the rms scatter of the difference between
the observed spectrum and synthetic fit is 0.010 and 0.027 for
the Sun and Arcturus, respectively.   Without the astrophysical 
values the rms scatter of the differences were 0.017 and 0.032
for the Sun and Arcturus, respectively.  The improvement is 
smaller for Arcturus since the spectrum and residuals are dominated
by molecular features which are not adjusted in our
astrophysical methodology for DR12.   The evidence of how
the line list performs is in its ability to derive quantitative abundances
which can be compared to literature values for the same stars.
\citet{Holtzman2015} shows that over the range: 
$T_{eff} = 3800 - 5300$ K, log $g = 0.5 - 3.8$ dex, [M/H] $= -2 - +0.5$ dex,  ASPCAP
delivers $T_{eff}$ to within 91.5 K, log $g$ to within 0.3 dex and 
absolute abundances of 14 elements of $0.1 - 0.2$ dex accuracy.  
This scatter is in part due to errors in the line list, but also 
includes errors in carbon isotope ratio mismatch, line spread function 
modeling errors, NLTE affects, and lack of spherical model atmospheres.

\section{ Missing Lines \label{missing}}

\begin{figure}
\includegraphics[scale=.50,angle=-90]{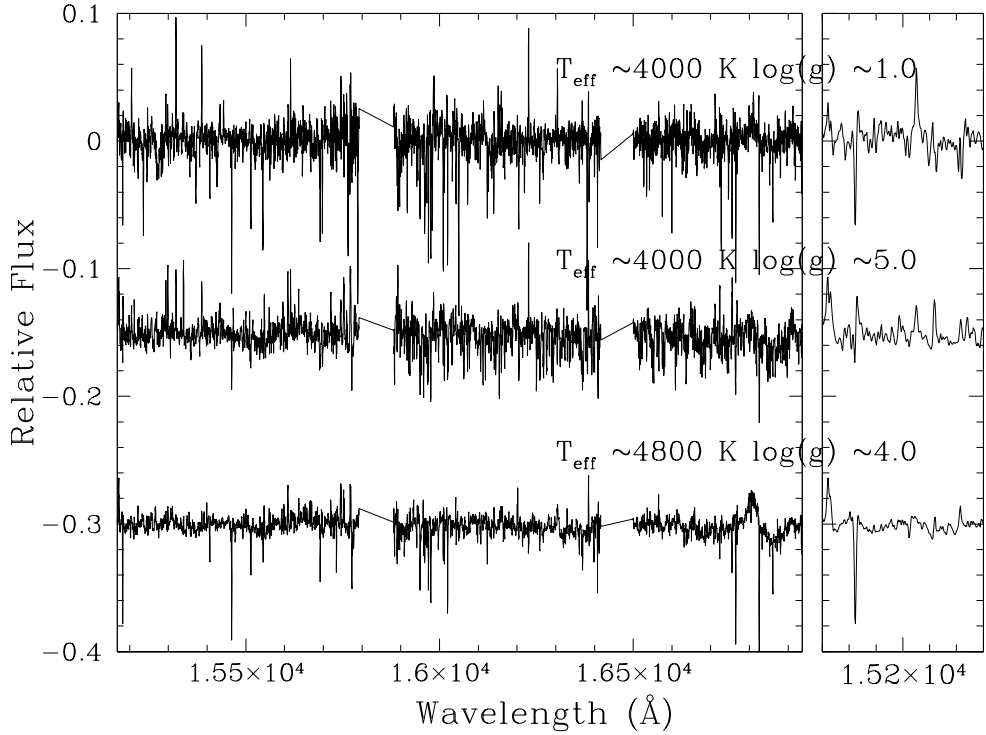}
\caption{The average residual spectra (observed spectra - best fit synthetic spectra) for stars
near $T_{eff} \sim$ 4000 K and log($g$) $\sim$ 1.0 cgs dex (top),  $T_{eff} \sim$4000 K and 
log $g\sim$5.0 cgs dex (middle), and $T_{eff} \sim$4800 K and log g$\sim$4.0 cgs dex (bottom).  
The left pane shows the entire APOGEE region; the breaks in the spectrum are the locations of the 
gaps between the detectors.  The right pane contains a small region of these average residual 
spectra.
\label{residual}
}
\end{figure}

The analysis of more than a hundred thousand stars in APOGEE allows us to determine where the line list is either 
missing features or has inaccurate log($gf$) values.  
To construct the list of missing lines we start by using the 
average difference
between the best fit synthetic spectra and the observed spectra shifted to the rest wavelength for stars of specific 
effective temperatures and gravities.  
We have chosen to construct three average residual spectra, the first from all 
giant stars with  $T_{eff} \sim$4000 K and log $g\sim$1.0 cgs dex, 
the second from all dwarf stars with T$_{eff} \sim$4800 K and log g$\sim$4.0 cgs dex, and the last one from all 
dwarf stars with  $T_{eff} \sim$4000 K and log $g\sim$5.0 cgs dex.      These three residual spectra are shown in 
Figure \ref{residual}.   The left pane shows the entire spectra including the gaps between the APOGEE detectors and
the right pane shows a small portion of these spectra where a few strong features and several weaker features can be seen 
in more detail.  
We tabulated all residual features with depth greater than 1\% (Table \ref{Missing Lines}).   
For completeness we also included the unidentified features in the Be stars observed by APOGEE as telluric calibrators noted by \citet{Chojnowski14}.
We supplemented this list with the strong lines that were not 
properly modeled in the analysis of a set of  Fourier Transform Spectrograph (FTS) spectra of standard stars in \citet{Smith13}.   
Some of these lines fall between the detectors
of the APOGEE instrument.    For the coolest dwarfs we suspect that some of the residual features may be H$_2$O and FeH, since those
lines are known to be detected in M dwarfs in the $H$-band.   As mentioned in Section \ref{FeH}, we were unable to 
find an FeH line list that produced significant improvements for the few M dwarfs we tested.   
Future work on dwarfs includes MARCS/Turbospectrum modeling on these stars by including H$_2$O and FeH  \citep{Zamora15}.

In principle, shifting the spectra to their rest wavelengths and averaging in that frame 
should have eliminated residual features such as badly removed telluric emission and absorption features,
as well as any interstellar features.   However, due to the complicated selection biases in APOGEE targeting
it is possible that some interstellar or telluric features populate Table \ref{Missing Lines}.  One example is
the feature at 15272.4 \AA\,  which is the strongest diffuse interstellar band (DIB) 
 in the $H$-band and is detected in the majority of APOGEE spectra \citep{Zasowski15}.  
We hope that future versions of the APOGEE line list will include identifications of these types of features.

\begin{table} 
\begin{center}
\caption{
\label{Missing Lines}
Missing Lines}
\begin{tabular}{|l|r|r|}
\tableline\tableline
 $\lambda$ (vac \AA) &  EW (m\AA) &  notes \\
\tableline
15174.1 & 18 & a \\
15177.5 & 12 & h \\
15178.6 & 25 & a \\
15180.4 & 20 & a \\
15182.3 & 61 & a, b, h \\
15187.3 & 19 & a \\
15201.1 & 16 & a \\
15202.4 & 24 & a, b, h \\
15209.7 & 39 & a \\
15210.5 & 17 & h \\
15211.0 & 33 & b \\
15212.5 & 48 & a \\
15214.5 & 16 & a \\
15216.3 & 20 & a, b \\
15219.4 & 59 & b \\
15227.6 & 10 & h \\
15228.7 & 12 & h \\
 \tableline   
 \end{tabular}
 \tablecomments{Table \ref{Missing Lines}  is published in its entirety in the
electronic edition of the {\it Astrophysical Journal Supplements}.  A portion is
shown here for guidance regarding its form and content.
}
\tablenotetext{a}{Seen in the cool APOGEE giant spectra residual }
\tablenotetext{b}{Seen in the hot APOGEE spectra residual }
\tablenotetext{c}{Between APOGEE detectors }
\tablenotetext{d}{all FTS stars }
\tablenotetext{e}{only M FTS stars }
\tablenotetext{f}{found only in HD199799 }
\tablenotetext{g}{see S. Hasselquist et al. 2016 in prep. }
\tablenotetext{h}{Seen in cool APOGEE dwarf spectra residuals}
\tablenotetext{i}{Seen in APOGEE Be spectra, see \citet{Chojnowski14}}
\tablenotetext{j}{This may be a DIB feature, see \citet{Zasowski15}}
\tablenotetext{k}{This may be a DIB feature, see \citet{Geballe11}}
\end{center}
\end{table}

\section{Summary \label{formats}}

The machine readable tables and data files contain the DR10, INT and 
DR12 line lists in the ASPCAP format, see Tables \ref{FormatDR12} -
\ref{FormatDR10}.  Additional copies formatted for
MOOG \citep{MOOG}, Synspec \citep[][]{Hubeny2006, Hubeny2011}  and Turbospectrum
\citep{Turbo98,Plez2012} are available in the electronic edition for download.
The ASPCAP formatted line lists contain the baseline line list, as well as the astrophysical log($gf$) and damping values.  This
line list is formatted in a way similar to that employed by
Kurucz in his online database and CDROM releases, where wavelengths are given vacuum and in nm.
Note that over the wide range of this line list, the choice of what 
methodology to use to translate from air to vacuum can
have a significant impact.   We have adopted the conversions of \citet{C96}.   See Appendix \ref{AirVac} for further discussion
on this choice and coefficients for conversion.
For the upper and lower energy levels we use the unit $cm^{-1}$.   

Some attention has been paid to HFS within these line lists but in the very strongest lines of Al the lack
of the HFS components means that the abundances derived are likely to be over-estimated.   Users of this line list and the 
abundances of Al should be wary in the strongest lined and coolest stars.

This line list has been tuned to be applied to red giants over a range that
covers F,G,K and warm M giants.   Tests conducted by the APOGEE team suggest that
it also performs adequately for F, G, and K dwarfs.  These line lists does not 
contain the FeH lines nor the H$_2$O lines needed to model cool M giants or M dwarfs.   
The reader is cautioned when using these line lists outside these stellar types or on stars with extreme abundance
patterns such as carbon stars where other molecular features may play an important role and these line lists have
not been tested.   
In addition, the use of the astrophysical log($gf$) values
derived for this work should be approached with caution when using a different
spectral synthesis code or suite of model atmospheres.  See Sections 3 and 4 of
\citet{Zamora15} for more extensive information about the tests that
have been conducted and possible systematics that could be introduced.

The ASPCAP machine readable line lists (Tables \ref{FormatDR12} -
\ref{FormatDR10}), are $\sim24$ MB in size with 35 columns and $\sim
134,000$ rows of data.



\begin{table} 
\begin{center}
\caption{
\label{FormatDR12}
ASPCAP Line List Format for DR12 (20131216)}
\begin{tabular}{|l|c|r|l|}
\tableline\tableline
 Bytes &  Format &  Label & Description \\
\tableline
  1-  9  & F9.4  &  lam   &         wavelength in vacuum nm \\
 11- 17  & F7.3   & orggf   &         original log($gf$) value \\
 19- 25  & F7.3  &  newgf   &         improved literature or laboratory log($gf$) \\
 27- 30  & F4.2  &  enewgf  &         error on log($gf$), when available \\
 32- 34  & A3     & snewgf      &        source for improved literature or laboratory log(gf) \\
 35- 41  & F7.3  &  astgf   &         astrophysical log($gf$) \\
 43- 45  & A3     & sastgf     &         source for astrophysical log($gf$) \\
 47- 54  & F8.2   & specid     &         species id \\
 55- 66  & F12.3  & EP1     &      lower Energy Level in cm$^{-1}$ \\ 
 67- 71  & F5.1   & J1            &      $J$ value for EP1 \\
 72- 82  & A11    & EP1id        &       EP1 level identification \\
 83- 94  & F12.3  & EP2     &       upper Energy Level in cm$^{-1}$ \\
 95- 99  & F5.1   & J2            &      $J$ value for EP2 \\
100-110  & A11    & EP2id     &          EP2 level identification \\
111-116  & F6.2   & Rad          &       Damping Rad \\
117-122  & F6.2   & Sta            &     Damping Stark \\
123-128  & F6.2   & vdW         &        Damping vdW \\
130-131  & I2     & unlte          &     NLTE level number upper \\
132-133  & I2     & lnlte          &     NLTE level number lower \\
134-136  & I3     & iso1          &      isotope number \\
137-142  & F6.3   & hyp         &        hyperfine component log fractional strength \\
143-145  & I3      &iso2          &      isotope number \\
146-151  & F6.3  &  isof         &       log isotopic abundance fraction \\ 
152-156  & I5     & hE1     &      hyperfine shift for first level to be added to E1 in mK \\
157-161  & I5    &  hE2    &         hyperfine shift for first level to be added to E2 in mK \\
162      & A1     & F0          &        hyperfine F symbol \\
163      & I1     & F1          &        hyperfine F for the first level \\
164      & A1     & note1   &            note on character of hyperfine data for first level: $z$ none, ? guessed \\
165      & A1     & S          &         the symbol "--" for legibility \\
166      & I1     & F2            &      hyperfine F' for the second level \\
167      & A1     & note2       &        note on character of hyperfine data for second level: $z$ none, ? guessed \\
168-172  & I5     & g1           &       lande $g$ for first level times 1000 \\
173-177  & I5    &  g2           &       lande $g$ for second level times 1000 \\
178-180  & A3     & vdWorg    &          source for the original vdW damping \\
181-186  & F6.2  &  vdWast      &        astrophysical vdW damping \\
\tableline   
\end{tabular}
\end{center}
\tablecomments{\\In addition the same DR data is available in MOOG, Turbospectrum and
ASS{$\epsilon$}T formats in the included .tar.gz package. \\
This table is available in its entirety in FITS format. \\
This table is available in its entirety in machine-readable form. \\}
\end{table}

\clearpage

\begin{table}
\begin{center}
\caption{
\label{FormatINT}
ASPCAP Line List Format for INT (20120216)}
\begin{tabular}{|l|c|r|l|}
\tableline\tableline
 Bytes &  Format &  Label & Description \\
\tableline
\multicolumn{4}{c}{Same Format as Table \ref{FormatDR12} }\\
\tableline  
\end{tabular}
\end{center}
\tablecomments{\\In addition the same DR data is available in MOOG, Turbospectrum and
ASS{$\epsilon$}T formats in the included .tar.gz package. \\
This table is available in its entirety in FITS format. \\
This table is available in its entirety in machine-readable form. \\}

\end{table}

\clearpage

\begin{table}
\begin{center}
\caption{
\label{FormatDR10}
ASPCAP Line List Format for DR10 (20110510)}
\begin{tabular}{|l|c|r|l|}
\tableline\tableline
 Bytes &  Format &  Label & Description \\
\tableline
\multicolumn{4}{c}{Same Format as Table \ref{FormatDR12} }\\
\tableline  
\end{tabular}
\end{center}
\tablecomments{\\In addition the same DR data is available in MOOG, Turbospectrum and
ASS{$\epsilon$}T formats in the included .tar.gz package. \\
This table is available in its entirety in FITS format. \\
This table is available in its entirety in machine-readable form. \\}

\end{table}

\clearpage

For a detailed discussion on how these line lists performed we point the reader to \citet{Meszaros13} and \citet{Lamb15}
for  DR10 performance,   \citet{Smith13}  for discussions
of the INT line list and \citet{Meszaros15} and \citet{Holtzman2015} for DR12 performance.   
 \citet{Meszaros13} compared literature metallicity, effective temperature, and surface gravity to the values computed by ASPCAP for
 DR10.    They found some systematic offsets in gravity and metallicity.     The source of these offsets may not be the line
 list but rather the methodology within the DR10 version of ASPCAP.  
 \citet{Lamb15}  conducted an independent
manual analysis of several metal-poor globular clusters stars with optical spectra and APOGEE DR10 spectra and line list.   
They found good agreement for Fe, Mg, and Ca lines, but had a concern about the single detectable Ti I line (15339 \AA\ vac) 
being a blend.   \citet{Lamb15} adopted an average between the optical and $H$-band results for Fe, Mg, and Ca and 
adopted the $H$-band results for Al, Si, O, C, and N.
 \citet{Smith13}  conducted an independent manual analysis of several very high resolution, very high S/N FTS spectra using 
the INT line list.  In a comparison
of their results to the literature, they found agreement within $\sim$0.1 dex for all abundances they derived:
$^{12}$C, $^{13}$C, N, O, Mg, Al, Si, K, Ca, Ti, V, Cr, Mn, Fe, Co, Ni, and Cu.   
\citet{Cunha2015} analyzed manually, also using the INT line list,  
a sample of red giants and clump stars in the very metal rich open cluster NGC6791 and
found overall good agreement with optical results from the literature for the 
abundances of oxygen, sodium and iron.

\citet{Meszaros15}  derived 
abundances in an independent manual analysis of APOGEE spectra of 10 globular clusters using the DR12 line list.   
They derived abundances for nine elements and found good agreement for most elements, once zero point differences in 
adopted solar abundances are considered.   Exceptions included Ti and Ca, which exhibited a large scatter in the APOGEE results, and
Al, which are based on very weak spectral lines in the literature optical analyses and on very strong features in the $H$-band.      \citet{Holtzman2015}
analyzed the ASPCAP DR12 results for self consistency within clusters and also did comparisons with abundance results in
the literature.  They found that DR12 results had an
internal abundance consistency at the level of 0.05 -- 0.09 dex and 0.1 -- 0.2 dex agreement with literature values.   
 This study, however, pointed out a number of elements  which exhibited unexpected trends with respect to metallicity. 
 In particular, for the entire APOGEE sample, the mean abundances of S, Si, and Ca at roughly solar metallicities were
above the solar value, 
and Ti does not exhibit the expected rise at decreasing metallicity seen in the literature.     These types of analyses and, in 
particular, comparisons with cluster abundance results can act as a guide for 
future improvements to the line list.  

\appendix 

\section{Air-Vacuum Conversion \label{AirVac}}

Since the APOGEE spectrograph is operating in vacuum, it was natural for APOGEE to adopt a vacuum wavelength scale.
Further support for embracing a vacuum scale is motivated by the fact that, historically, the definition of standard air includes tight constrains on temperature (15 C), pressure (1 atmosphere), and humidity (dry), but not so clearly the CO2 concentration, which is time and spatially dependent, and the standard temperature scale has also seen multiple changes in the past century, and these drag the air-to-vacuum corrections. In addition, the reference wavelengths for the Th and Ar lines usually employed for calibration come from vacuum-housed spectrographs, and are therefore more naturally used in vacuum \citep{Norlen1973, Palmer1983, Hinkle2001, Lovis2007, Kerber2008}.
This appendix describes the differences on the vacuum-to-air conversions and motivates APOGEE adopting a specific one for dealing with such corrections.

\subsection{Available formulae}

The IAU standard for the vacuum to standard air corrections (see resolution No. C15, Commission 44, XXI General Assembly in 1991) refers to \citet{Oosterhoff1957}, which adopts the results by \citet{Edlen1953}.

\begin{equation}\label{vac}
\frac{\lambda_0 -\lambda}{\lambda} = n - 1 = a + \frac{b1}{c1 - 1/\lambda^2_0} + \frac{b2}{c2 - 1/\lambda^2_0}
\end{equation}
where $\lambda_0$ is given in $\mu$m, and the constants are given in the first row of Table \ref{vacvalues}.
Later work widely adopted in the physics literature by \citet{Edlen1966} rederived the constants from optical and near UV data, and later \citet[and references therein]{Peck1972} added additional measurements extending into the IR, up to 1.7 $\mu$m,
which showed a systematic deviation from Edl{\'e}nÕs equation at the level of several 10$^{-9}$, or a few ms$^{-1}$.

\begin{table} 
\begin{center}
\caption{Parameters for Eq. \ref{vac}
\label{vacvalues}}
\begin{tabular}{|c|ccccc|}
\tableline\tableline
Reference &  a & b1 & b2 & c1 & c2 \\
\tableline
\citet{Edlen1953} & 6.4328$\times10^{-5}$ & 2.94981$\times10^{-2}$ & 2.5540$\times10^{-4}$ & 146.0 & 41.0 \\
\citet{Edlen1966} & 6.4328$\times10^{-5}$ & 2.406030$\times10^{-2}$ & 1.5997$\times10^{-4}$ & 130.0 & 38.9 \\
\citet{Peck1972} & 0.0 & 5.791817$\times10^{-2}$ & 1.67909$\times10^{-3}$ & 238.0185 & 57.362 \\
\citet{C96} & 0.0 & 5.792105$\times10^{-2}$ & 1.67917$\times10^{-3}$ & 238.0185 & 57.362 \\
\tableline   
\end{tabular}
\end{center}
\end{table}


\begin{figure}
\epsscale{0.8}
\plotone{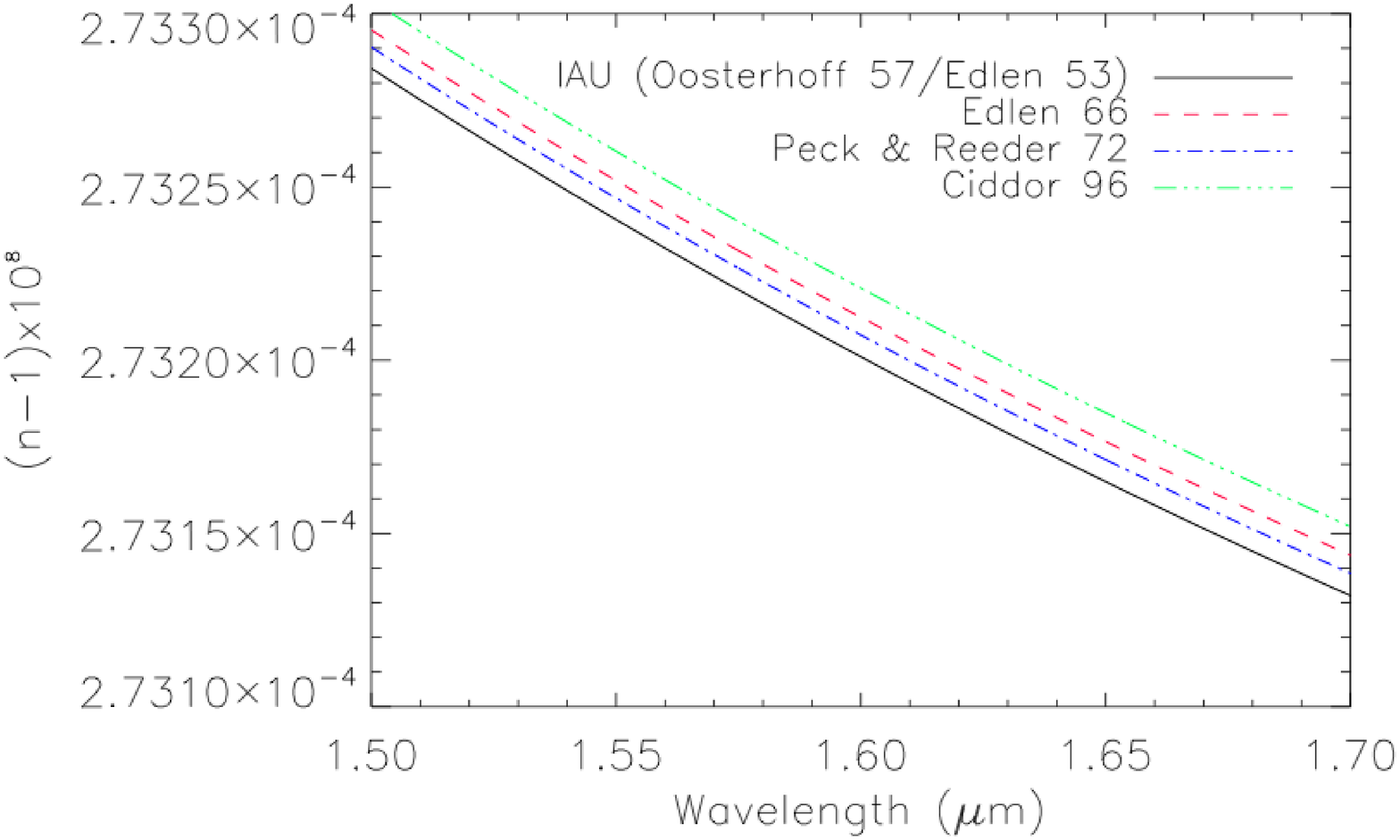}
\caption{
Difference between the refractive index of standard air ($n$) and unity in the $H$-band for four different sources considered.
\label{vaccuum2}
}
\end{figure}

Table \ref{vacvalues} compares the parameters proposed by \citet{Edlen1953} \citet{Edlen1966}, \citet{Peck1972}, and \citet{C96}, the 
latter reference largely based on \citet{Peck1972}, but updated to account for the changes taken place in the international 
temperature scale since their work, and adjusting the results for the CO$_2$ concentration.
Fig. \ref{vaccuum2} shows a maximum discrepancies at the level of 3 $\times$ 10$^{-8}$, or 10 ms$^{-1}$ at 1.6 $\mu$m. 
The equations proposed by \citet{Edlen1966} and \citet{Peck1972} differ only by about one fourth of that. About half 
of the difference between Peck \& Reeder and Ciddor (2 $\times$ 10$^{-8}$ or 6.5 ms$^{-1}$ at 1.6 $\mu$m) is related to the temperature scale. 
The paper by Peck \& Reeder is not specific regarding the scale used, and Ciddor assumes they used the IPTS-48 standard\footnote{
International Practical Temperature Scale; $t_{68} = 1.00024 \times t_{90}$ \citep{Saunders1990}; 
$t_{68} = t_{48} - 4.4 \times 10^{-6}t_{48} \times (100 - t_{48})$ \citep{Fofonoff1975}}, 
which is warmer than the one currently in use (ITS-90) by 9.2 mK at 15 C. If Peck \& Reeder used the IPTS-68 standard instead, 
the difference would be reduced to 5.6 mK. Changes of 9.2 mK and 5.6 mK amount to an increase in ($n -  1$) $\times  c$ of about 2.6 
and 1.6 m$^{-1}$, respectively, at 1.6  $\mu$m. The rest of the correction between Peck \& Reeder and CiddorÕs review (3.9 ms$^{-1}$ 
at 1.6 $\mu$m) is related to the CO$_2$ concentration in standard air. Because of secular variations in the typical laboratory 
air, Birch \& Downs proposed to use 450 ppm in the definition of standard air, while Ciddor
estimates that a value closer to 300 ppm is adequate for the measurements used by Peck \& Reeder. In summary, the 
variation between between \citet{Peck1972} and \citet{C96} is connected to a change between the actual conditions 
of standard air now and at the time of the measurements.

\subsection{Conclusions}

In view of the preceding discussion, we underline the advantages of using vacuum wavelengths for the APOGEE spectra. In some cases, corrections between air and vacuum will be needed, and for those it is proposed that the formulation proposed by \citet{C96} be used. This corresponds to Eq. \ref{vac} with the Ciddor constants given in Table \ref{vacvalues}. This is valid for the wavelength range between 0.23 and 1.7 $\mu$m, and the estimated accuracy in the predicted refraction index of standard air is about 10$^{-8}$ or roughly 3 ms$^{-1}$ at 1.6 $\mu$m. Note that it is straightforward to go from vacuum to standard air wavelengths with Eq. \ref{vac}, but the inverse process requires iteration since n is given as a function of vacuum wavelength.

\acknowledgments{

Funding for SDSS-III has been provided by the Alfred P. Sloan Foundation, the Participating Institutions, the National Science Foundation, and the U.S. Department of Energy Office of Science. The SDSS-III web site is \url{http://www.sdss3.org/}.

SDSS-III is managed by the Astrophysical Research Consortium for the Participating Institutions of the SDSS-III Collaboration including the University of Arizona, the Brazilian Participation Group, Brookhaven National Laboratory, Carnegie Mellon University, University of Florida, the French Participation Group, the German Participation Group, Harvard University, the Instituto de Astrofisica de Canarias, the Michigan State/Notre Dame/JINA Participation Group, Johns Hopkins University, Lawrence Berkeley National Laboratory, Max Planck Institute for Astrophysics, Max Planck Institute for Extraterrestrial Physics, New Mexico State University, New York University, Ohio State University, Pennsylvania State University, University of Portsmouth, Princeton University, the Spanish Participation Group, University of Tokyo, University of Utah, Vanderbilt University, University of Virginia, University of Washington, and Yale University.

S. M. has been supported by the J{\'a}nos Bolyai Research
Scholarship of the Hungarian Academy of Sciences.
D. A. G. H. and O. Z. acknowledge support provided by the Spanish Ministry of Economy
and Competitiveness under grants AYA-2011-27754 and AYA-2014-58082-P.
DB was supported by grant RSF 14-50-00043.
We would like to 
thank R. L. Kurucz for his many fundamental contributions that have made this work possible.  

}

@MISC{2015ascl.soft02022B,
    author = {{Bizyaev}, D. and {Shetrone}, M.},
    title = "{AstroLines: Astrophysical line list generator in the $H$-band}",
    howpublished = {Astrophysics Source Code Library},
    year = 2015,
    month = feb,
    archivePrefix = "ascl",
    eprint = {1502.022},
    adsurl = {http://adsabs.harvard.edu/abs/2015ascl.soft02022B},
    adsnote = {Provided by the SAO/NASA Astrophysics Data System}
} 

\bibitem[ Blackwell-Whitehead et al.(2006)] {BLNPJLPV06}  Blackwell-Whitehead, R.J., Lundberg, H., Nave, G. ,  Pickering, J. C., Jones, H. R. A. , Lyubchik, Y. , Pavlenko, Y. V. , \&  Viti, S. 2006, Mon. Not. R. Astron. Soc. 373, 1603Ð1609 (2006)

\bibitem[Brault et  al.(1982)]{Brault} Brault, J.~W., Testerman, L., Grevesse, N., et al.\ 1982, \aap, 108, 201

\bibitem[Castelli et al.(1997)]{Castelli97} Castelli, F., Gratton, R.~G., \& Kurucz, R.~L.\ 1997, \aap, 318, 841

\bibitem[Chojnowski et al.(2014)]{Chojnowski14} Chojnowski, S.~D., Whelan, D.~G., Wisniewski, J.~P., et al.\ 2014, arXiv:1409.4668 

\bibitem[Ciddor(1996)] {C96} Ciddor, P. E. 1996, Applied Optics, 35, 1566

\bibitem[Claret(2000)]{claret00} Claret, A. 2000, \aap, 363, 1081

\bibitem[Cunha et al.(2015)]{Cunha2015} Cunha, K., Smith, V. V., Johnson, J. A., et al. 2015, \apjl , 798, L41

\bibitem[Dulick et al.(2003)]{Dulick} Dulick, M., Bauschlicher, C.~W., Jr., Burrows, A., et al.\ 2003, \apj, 594, 651

\bibitem[Eisenstein et al.(2011)]{Eisenstein11} Eisenstein, D.J., Weinberg, D.H., Agol, E., Aihara, H., Allende Prieto, C., Anderson, S.F., Arns, J.A., Aubourg, E., Bailey, S., Balbinot, E. and 234 coauthors 2011, AJ, 142, 72

\bibitem[Edl{\'e}n(1953)]{Edlen1953} Edl{'e}n B., 1953, J. Opt. Soc. Am., 43, 339

\bibitem[Edl{\'e}n(1966)]{Edlen1966} Edl{'e}n B., 1966, Metrologia, 2, 71

\bibitem[Falkenburg \& Zimmermann(1979)]{Falkenburg79} Falkenburg \& Zimmermann 1979 , Naturforsch 34a, 1249

\bibitem[Fofonoff \& Bryden(1975)]{Fofonoff1975} Fofonoff, N. P., \& Bryden, H. 1975, Journal of Marine Research, 33, Supple- ment, pp 69-82.

\bibitem[Geballe et al.(2011)]{Geballe11} Geballe, T.~R., Najarro, F., Figer, D.~F., Schlegelmilch, B.~W., \& de La Fuente, D.\ 2011, \nat, 479, 200 

\bibitem[Garc{\'{\i}}a P{\'e}rez et al.(2015)]{Garcia2015}
Garc{\'{\i}}a P{\'e}rez, A.~E., Allende Prieto, C., Holtzman, J.~A., et
al.\ 2015, arXiv:1510.07635

\bibitem[Garc\'{\i}a P\'erez et al.(2013)]{GarciaPerez13} Garc\'{\i}a P\'erez, A.E., Cunha, K., Shetrone, M., Majewski, S.R., Johnson, J.A., Smith, V.V., Schiavon, R.P., Holtzman, J., Nidever, D., Zasowski, G., and 20 coauthors  2013, ApJL 767, 9.

\bibitem[Goldman et al.(1998)]{Goldman} Goldman, A., Shoenfeld, W. G., Goorvitch, D., Chackerian, C., Jr., Dothe, H., Melen, F., Abrams, M. C., \& Selby, J. E. A. 1998, J. Quant. Spectrosc. Radiat. Transfer, 59, 453

\bibitem[Goorvitch(1994)]{Goorvitch} Goorvitch, D.\ 1994, \apjs, 95, 535

\bibitem[Gray(1981)]{Gray1981} Gray, D.F. 1981, \apj 245, 992

\bibitem[Gray \& Brown(2006)]{Gray2006} Gray, D.F, \& Brown, K.I. 2006 \pasp 118 1112

\bibitem[Grevesse et al.(1996)]{Grevesse96} Grevesse, N., Noels,
A., \& Sauval, A.~J.\ 1996, Cosmic Abundances, 99, 117 

\bibitem[Gustafsson et al.(2003)]{Gustafsson03} Gustafsson, B.,
Edvardsson, B., Eriksson, K., et al.\ 2003, Stellar Atmosphere Modeling,
288, 331

\bibitem[Gustafsson et al.(2008)]{Gustafsson08} Gustafsson, B., Edvardsson, B., Eriksson, K., Jorgensen, U. G., Nordlund, A., \& Plez, B. 2008, AA, 486, 951

\bibitem[Gunn et al.(2006)]{gunn01} Gunn, J.~E., Siegmund, W.~A., Mannery, E.~J. et al. 2006, AJ, 131, 2332

\bibitem[G{\"u}zel{\c c}imen et al.(2011)]{Guzelcimen} G{\"u}zel{\c c}imen Ba{\c s}ar, G., {\"O}zt{\"u}rk, I.~K., et al. 2011, Journal of Physics B Atomic Molecular Physics, 44, 215001.

\bibitem[G{\"u}zel{\c c}imen et al.(2014)]{Guzelcimen14} G{\"u}zel{\c c}imen, F., Yap{\i}c{\i}, B., Demir, G., et al.\ 2014, \apjs, 
214, 9 

\bibitem[Hansen et al.(1999)]{Hansen99} Hansen, Laughlin, van der Hart, Verbockhaven 1999, J. Phys B: At. Mol. Opt. Phys. 32 2099

\bibitem[Happer(1974)]{Happer74} Happer, W. in Atomic Physics 4, edited by G. zu Putlitz, E.W. Weber, \& A. Winnacker (Plenum Press, New York, 1974) pp. 651Ð682.

\bibitem[Hinkle et al.(2001)]{Hinkle2001} Hinkle, K.~H., Joyce, R.~R., Hedden, A., Wallace, L., \& Engleman, R., Jr.\ 2001, \pasp, 113, 548

\bibitem[Hinkle et al.(1995)] {Hinkle95} Hinkle, K., Wallace, L., Livingston, W., D. 1995, \pasp 107, 1042.

\bibitem[Holtzman et al.(2015)]{Holtzman2015} Holtzman, J.A., Shetrone, M., Johnson, J.A. et al. 2015, arXiv:1501.04110

\bibitem[Hubeny \& Lanz(2011)]{Hubeny2011} Hubeny, I., \& Lanz, T.\ 2011, Astrophysics Source Code Library, 1109.022

\bibitem[Hubeny(2006)]{Hubeny2006} Hubeny, I.\ 2006, Computational Methods in Transport, 15

\bibitem[Kellerher \& Podobedova(2008)]{KP08} Kelleher, D.E. \&  Podobedova L.I. 2008, J. Phys. Chem. Ref. Data 37, 267 

\bibitem[Kerber et al.(2008)]{Kerber2008} Kerber, F., Nave, G., \& Sansonetti, C. J. 2008, ApJS, 178, 374

\bibitem[Koesterke(2009)]{Koesterke09} Koesterke, L.\ 2009,
American Institute of Physics Conference Series, 1171, 73

\bibitem[Kokkin et al.(2007)]{Kokkin}  Kokkin D.~L., Bacskay G~B., Schmidt T.~W 2007 (JournalChemPhys, v126, 084302).

\bibitem[Kramida et al.(2014)]{Kramida14} Kramida, A., Yu. Ralchenko, Reader, J., \& and NIST ASD Team. 2014, NIST Atomic Spectra Database (ver. 5.2), [Online]. Available: \url{http://physics.nist.gov/asd} [2014, October 17]. National Institute of Standards and Technology, Gaithersburg, MD.

\bibitem[Lamb et al.(2015)]{Lamb15} Lamb, M., Venn, K., Shetrone, M., Sakari, C., \& Pritzl, B. 2015, \mnras, 448, 42L.

\bibitem[Langhoff  \& Bauschlicher(1990)]{Langhoff} Langhoff, S.~R., \& Bauschlicher, C.~W.\ 1990, Journal of Molecular Spectroscopy, 141, 243 

\bibitem[Laughlin(1992)]{Laughlin} Laughlin, C. 1992 , Physica Scripta 45, 238.

\bibitem[Lawler et al.(2013)]{LGWSC13} Lawler, J.~E., Guzman,
A., Wood, M.~P., Sneden, C., \& Cowan, J.~J.\ 2013, \apjs, 205, 11

\bibitem[Livingston \& Wallace(1991)]{LW91} Livingston, W., \& Wallace, L.\ 1991, NSO Technical Report, Tucson: National Solar Observatory, National Optical Astronomy Observatory, 1991

\bibitem[Lovis \& Pepe(2007)]{Lovis2007} Lovis, C., \& Pepe, F. 2007, \aa, 468, 1115

\bibitem[Majewski et al.(2015)]{Majewski15} Majewski, S.~R., Schiavon, P., Frinchaboy, P.M. et al.\ 2015, arXiv:1509.05420

\bibitem[Martin, Fuhr, \& Wiese(1988)]{MFW88} Martin, G.A., Fuhr, J. R.  \& Wiese, W. L.  1988, J. Phys. Chem. Ref. Data 17, Suppl. 3, 512 

\bibitem[McWilliam et al.(1995)]{McWilliam95}  McWilliam, A., Preston, G.W., Sneden, C., \& Searle, L. 1995, \aj, 109, 2757

\bibitem[Melendez \& Barbuy(1999)]{MB99} Melendez, J. \& Barbuy, B. 1999, ApJS 124, 527

\bibitem[M{\'e}sz{\'a}ros et al.(2012)]{Meszaros12} M{\'e}sz{\'a}ros, S., Allende Prieto, C., Edvardsson, B., et al.\ 2012, \aj, 144, 120

\bibitem[M{\'e}sz{\'a}ros et al.(2013)]{Meszaros13} M{\'e}sz{\'a}ros, S., Holtzman, J., Garc{\'{\i}}a P{\'e}rez, A.~E., et al.\ 2013, \aj, 146, 133 

\bibitem[M{\'e}sz{\'a}ros et al.(2015)]{Meszaros15}  M{\'e}sz{\'a}ros, S., Martell, S.~L., Shetrone, M., et al.\ 2015, \aj, 149, 153

\bibitem[Nelder \& Mead(1965)]{NM65} Nelder, J.A., \& Mead, R. 1965,  Computer Journal 7: 308Ð313

\bibitem[Nidever et al.(2015)]{Nidever15} Nidever, D.~L., Holtzman, J.~A., Allende Prieto, C., et al.\ 2015, arXiv:1501.03742

\bibitem[Norl{\'e}n(1973)]{Norlen1973} Norl{\'e}n, G., 1973, Physica Scripta, 8, 24

\bibitem[Oosterhoff(1957)]{Oosterhoff1957} Oosterhoff, P. T. 1957, Trans. IAU, 9, 69 and 202

\bibitem[Palmer \& Engleman(1983)]{Palmer1983} Palmer, B.~A., \& Engleman, R.\ 1983, LA, Los Alamos: National Laboratory, |c1983,

\bibitem[Palmeri et al.(1995)]{Palmeri95} Palmeri, P., Biemont, E., Aboussaid, A., \& Godefroid, M.\ 1995, Journal of Physics B Atomic Molecular Physics, 28, 3741

\bibitem[Palmeri et al.(1997)]{Palmeri97} Palmeri, P., Biemont, E., P.Quinet, P., et al. 1997, Phys. Scripta 55, 586

\bibitem[Peck \& Reeder(1972)]{Peck1972} Peck, E. R., \& Reeder, K. 1972, J. Opt. Soc. Am., 62, 958

\bibitem[Plez(2012)]{Plez2012} Plez, B. 2012, Astrophysics Source Code Library, record ascl:1205.004

\bibitem[Podobedova, Kelleher \& Wiese(2009)]{PKW09} Podobedova, L.I., Kelleher, D.E. \& and Wiese, W.L. 2009,  J. Phys. Chem. Ref. Data 38, 171

\bibitem[Prato et al.(2002)]{Prato02} Prato, L., Simon, M., Mazeh, T., et al.\ 2002, \apj, 569, 863 

\bibitem[Ruffoni et al.(2013)]{RANP13} Ruffoni, M.~P., Allende
Prieto, C., Nave, G., \& Pickering, J.~C.\ 2013, \apj, 779, 17

\bibitem[Ryde et al. (2009)]{R09} Ryde, N., Edvardsson, B., Gustafsson, B., et al.\ 2009, \aap, 496, 701

\bibitem[Ryde et al.(2010)]{R10} Ryde, N., Gustafsson, B., Edvardsson, B., et al.\ 2010, \aap, 509, AA20 

\bibitem[Saloman(2012)]{S12} Saloman, E.B. 2012, JPCRD, 41, 013101

\bibitem[Safronova et al.(1999)]{Safronova} Safronova, M.~S., Johnson, W.~R., \& Derevianko, A.\ 1999, \pra, 60, 4476 

\bibitem[Sansonetti(2008)] {S08} Sansonetti, J.E. 2008,  J. Phys. Chem. Ref. Data 37, 1659

\bibitem[Sansonetti(2006)]{S06} J. E. Sansonetti, J.E. 2006, J. Phys. Chem. Ref. Data 35, 301;  Erratum: 37, 1183 (2008)

\bibitem[Saunders(1990)]{Saunders1990} Saunders, P. 1990. The International Temperature Scale of 1990, ITS-90. WOCE Newsletter 10.

\bibitem[Smith et al.(2013)]{Smith13} Smith, V.V., Cunha, K., Shetrone, M.D., M{\'e}sz{\'a}ros, S., Allende Prieto, C., Bizyaev, D., Garc\'{\i}a P\'erez, A., Majewski, S.R., Schiavon, R., Holtzman, J., Johnson, J.A. 2013, \apj 765, 16.

\bibitem[Sneden(1974)]{MOOG} Sneden, C. 1974, PhD Thesis, University of Texas at Austin.

\bibitem[Sneden et al.(2014)]{Sneden14} Sneden, C., Lucatello, S., Ram, R.~S., Brooke, J.~S.~A., \& Bernath, P.\ 2014, \apjs, 214, 26 

\bibitem[Sur et al.(2005)]{Sur2005} Sur, C., Chaudhuri, R.K., Das, B.P, \& Mukherjee, D. 2005, J Phys B 38, 4185

\bibitem[Thorne et al.(2011)]{T11} Thorne, A.~P.,
Pickering, J.~C., \& Semeniuk, J.\ 2011, \apjs, 192, 11

\bibitem[Unsold(1955)]{Unsold} Unsold, A.\ 1955, Berlin,
Springer, 1955.~2.~Aufl., Physik der Sternatmospharen, MIT besonderer Berucksichtigung der Sonne.

\bibitem[Wiese \& Fuhr(2007)]{WF07} W. L. Wiese W.L. \&  Fuhr J.R. 2007, J. Phys. Chem. Ref. Data 36, 1287; Erratum: 36, 1737 

\bibitem[Wiese \& Fuhr(2009)]{WF09} W. L. Wiese W.L. \&  Fuhr J.R. 2009, J. Phys. Chem. Ref. Data 38, 565; Erratum: 38, 1129

\bibitem[Wiese et al.(1996)]{WFD96} Wiese, W.~L., Fuhr, J.~R.,
\& Deters, T.~M.\ 1996, Atomic transition probabilities of carbon, nitrogen, and oxygen : a critical data compilation.~ Edited by W.L.~Wiese, J.R.~Fuhr, and T.M.~Deters.~Washington, DC :  American Chemical Society ...~for the National Institute of Standards and Technology (NIST) c1996.~QC 453 .W53 1996.~ Also Journal of Physical and Chemical Reference Data, Monograph 7.~ Melville, NY: AIP Press,

\bibitem[Wiese, Smith \& Miles(1969)]{WSM69}  Wiese, W.L., Smith, M.W., \& Miles B.M. 1969, in Nat. Stand. Ref. Data Ser., NSRDS-NBS 22, 268 pp. (U.S. Government Printing Office, Washington, D.C.)

\bibitem[Wyart \& Palmeri(1998)]{Wyart}  Wyart, J.-F., \& Palmeri, P. 1998,  \physscr, 58, 368

\bibitem[Wood et al.(2013)]{WLSC13} Wood, M.~P., Lawler,
J.~E., Sneden, C., \& Cowan, J.~J.\ 2013, \apjs, 208, 27

\bibitem[Wood et al.(2014)]{Wood14} Wood, M.~P., Lawler, J.~E., Den Hartog, E.~A., Sneden, C., \& Cowan, J.~J.\ 2014, \apjs, 214, 18

\bibitem[Zamora et al.(2015)]{Zamora15} Zamora, O., Garc{\'{\i}}a-Hern{\'a}ndez, D.~A., Allende Prieto, C., et al.\ 2015, \aj, 149, 181 

\bibitem[Zasowski et al.(2013)]{Zasowski13} Zasowski, G., Johnson, J.~A., Frinchaboy, P.~M., et al.\ 2013, \aj, 146, 81 

\bibitem[Zasowski et al.(2015)]{Zasowski15} Zasowski, G., M{\'e}nard,  B., Bizyaev, D., et al.\ 2015, \apj, 798, 35


\end{document}